\documentclass[aps,prx,twocolumn,citeautoscript,footinbib,eqsecnum,floatfix]{revtex4-1}
\usepackage[T1]{fontenc}
\usepackage{feynmp}
\usepackage{comment}
\usepackage{amsmath}
\usepackage{amssymb}
\usepackage{bbm}
\usepackage{braket}
\usepackage{xcolor}
\usepackage{pifont}
\usepackage{slashed}
\usepackage[mathscr]{euscript}
\usepackage[shortlabels]{enumitem}
\usepackage{tikz-feynman}
\usepackage[papersize={8.5in,11in}]{geometry}
\usepackage{standalone}
\allowdisplaybreaks

\usepackage{graphicx}
\usepackage[colorlinks=true]{hyperref}
\newcommand{\equref}[1]{Eq.~(\ref{#1})}
\newcommand{\equsref}[2]{Eqs.~(\ref{#1}) and (\ref{#2})}

\newcommand{\refcite}[1]{Ref.~\onlinecite{#1}}

\newcommand{\pdagger}{{\phantom{\dagger}}}
\newcommand{\diff}{\mathrm{d}}
\newcommand{\sign}{\,\text{sign}}
\renewcommand{\approx}{\simeq}

\newcommand{\vect}[1]{\boldsymbol{#1}}

\definecolor{wrongultramarine}{rgb}{1,0.5,0}

\hypersetup{
    bookmarks=true,         
    unicode=false,          
    pdftoolbar=true,        
    pdfmenubar=true,        
    pdffitwindow=false,     
    pdfstartview={FitH},    
    pdfsubject={},   
    pdfcreator={},   
    pdfproducer={}, 
    pdfkeywords={} {} {}, 
    pdfnewwindow=true,      
    colorlinks=true,       
    linkcolor=black, 
    citecolor=blue,        
    filecolor=magenta,      
    urlcolor=blue           
}

\newcommand{\beq}{\begin{equation}}
\newcommand{\eeq}{\end{equation}}
\def\bea{\begin{eqnarray}}
\def\eea{\end{eqnarray}}

\newcommand{\rd}{{\rm d}}
\newcommand{\sgn}{{\rm sgn\,}}

\begin{document}



\title{Gauge theories for the thermal Hall effect}

\author{Haoyu Guo}
\affiliation{Department of Physics, Harvard University, Cambridge MA 02138, USA}

\author{Rhine Samajdar}
\affiliation{Department of Physics, Harvard University, Cambridge MA 02138, USA}

\author{Mathias S. Scheurer}

\author{Subir Sachdev}
\affiliation{Department of Physics, Harvard University, Cambridge MA 02138, USA}

\date{\today}

\begin{abstract}
We consider the thermal Hall effect of fermionic matter coupled to emergent gauge fields in 2+1 dimensions. While the low-temperature thermal Hall conductivity of bulk topological phases can be connected to chiral edge states and a gravitational anomaly, there is no such interpretation at nonzero temperatures above 2+1 dimensional quantum critical points. In the limit of a large number of matter flavors, the leading contribution to the thermal Hall conductivity is that from the fermionic matter. The next-to-leading contribution is from the gauge fluctuations, and this has a sign which is opposite to that of the matter contribution. We illustrate this by computations on a Dirac Chern-Simons theory of the quantum phase transition in a square-lattice antiferromagnet involving the onset of semion topological order. We find similar results for a model of the pseudogap metal with Fermi pockets coupled to an emergent U(1) gauge field. We note connections to recent observations on the hole-doped cuprates: our theory captures the main trends, but the overall magnitude of the effect is smaller than that observed.
\end{abstract}

\maketitle
\tableofcontents

\section{Introduction}
\label{sec:intro}

Recent experiments have shown that the thermal Hall effect, also known as the the Righi-Leduc effect, is a powerful probe for the presence of unconventional excitations in correlated electron systems. For instance, in the spin liquid candidate $\alpha$-RuCl$_3$, the temperature and field dependence of the thermal Hall coefficient, $\kappa_{xy}$, has been suggested to indicate the presence of neutral excitations with exotic statistics \cite{2018PhRvL.120u7205K}.

Grissonnanche {\it et al.} \cite{Taillefer19} measured the thermal Hall effect in the normal state of four different copper-based superconductors. In the overdoped compounds, they observed a conventional $\kappa_{xy}$, related to the electrical Hall conductivity, $\sigma_{xy}$, by the Wiedemann-Franz law.
Interestingly, with decreasing doping, they observed the onset of a negative contribution to $\kappa_{xy}$ upon entering the pseudogap phase, which is unrelated to $\sigma_{xy}$.
This negative signal increases in magnitude with lowering doping, and persists all the way into the insulator. Given that a much smaller response is expected from conventional spin-wave theory \cite{SCSS19,han2019consideration}, Grissonnanche {\it et al.} argued that it indicated the presence of exotic neutral excitations in the pseudogap phase. In the present work, we employ a gauge theory for the pseudogap phase \cite{SS09,SCWFGS,WSCSGF,SS18,Scheurer_tri}, and propose that the emergent gauge field is a neutral excitation which could help produce the observed $\kappa_{xy}$.

In a previous study \cite{THNP}, we focused on the thermal Hall effect in the insulator, and illustrated that the proximity to a
quantum phase transition---between the N\'eel state and a state with coexisting N\'eel and semion topological order---could
explain the enhanced thermal Hall effect in the insulator. This enhanced $\kappa_{xy}$ was computed using the gauge theory for the critical point, which had four different formulations, all dual to each other. As the gauge theory is strongly coupled, the computation relied on an expansion in $1/N_f$, where $N_f$ is the number of flavors of matter fields. The calculation of $\kappa_{xy}$ at $N_f = \infty$ was described in \refcite{THNP}, and we will present further details here. We will also describe the structure of the leading $1/N_f$ corrections to $\kappa_{xy}$; we argue that an important component of these corrections (resulting from the analog of the `Aslamazov-Larkin' diagrams)
can be interpreted as the contribution of the collective mode associated with the emergent gauge field to $\kappa_{xy}$. This interpretation will be useful to us when we turn to consideration of the doped case in the latter part of this paper.

We now outline the models studied and the main results for the undoped and doped cases in turn.

\subsection{Undoped insulator}
\label{sec:undoped}

We focus our attention on one of the four duality-equivalent gauge theories describing the vicinity of the onset of semion topological order in the N\'eel state \cite{THNP}: the SU(2) gauge theory at Chern-Simons level $k$\,$=$\,$-1/2$, coupled to a single flavor ($N_f =1$) of a two-component Dirac fermion $\Psi$ with mass $m$. We generalize the fermions to $\Psi_\ell$ with $\ell$\,$=$\,$1 \ldots N_f$ flavors, and consider the Lagrangian
\begin{equation}
\label{eq:L_CS}
\mathcal{L}^{}_\Psi = \bar{\Psi}^\pdagger_\ell \left[i \gamma^\mu\left(\partial_\mu^{} - i A^{}_\mu \right) \right] \Psi^\pdagger_\ell + m\, \bar{\Psi}^\pdagger_\ell \Psi^\pdagger_\ell + k \, \mbox{\textit{CS}} [A^{}_\mu],
\end{equation}
where $A_\mu$ is the SU(2) gauge field, $k$ is the Chern-Simons level, and $\Psi$ is a continuum field derived from the lattice model in Appendix~\ref{LowEnergyFieldTheory}; $\bar{\Psi} = \Psi^\dagger \gamma^0$.
For $N_f = 1$ and $k=-1/2$, the $m>0$ phase of (\ref{eq:L_CS}) is `trivial' and describes the conventional N\'eel state,
while the $m<0$ phase of (\ref{eq:L_CS}) has semion topological order \cite{THNP}. The derivation of this field theory starting from a microscopic lattice model is sketched in Appendix~\ref{LowEnergyFieldTheory}.

The thermal Hall conductivity of (\ref{eq:L_CS}) is expected to obey
\beq
\kappa_{xy} = \frac{k_B^2 T}{\hbar} \mathcal{K}(m/T) \label{kappauniv}
\eeq
where $\mathcal{K}$ is
a dimensionless universal function of $m/T$, with $m$ the renormalized mass of the lowest quasiparticle excitations and $T$ the absolute temperature. In the limit $|m|/T \rightarrow \infty$, the exact values of $\kappa_{xy}$ can be deduced from arguments based on gravitational anomalies as \cite{Cappelli2001,THNP,Seiberg:2016}
\beq
\kappa^{}_{xy} = \frac{\pi k_B^2 T}{6 \hbar} \mbox{sgn}(\hat{k}) \left[ 2 |\hat{k}| - \frac{3|\hat{k}|}{|\hat{k}|+2}   \right]; \qquad \frac{|m|}{T} \rightarrow \infty,
\label{kappa_exact}
\eeq
where the integer $\hat{k}$ is defined by
\beq
\hat{k} = k + \frac{N_f}{2} \mbox{sgn}(m),
\eeq
and the sign function vanishes for zero argument, i.e., $\mbox{sgn}(0)=0$. We will obtain the large-$N_f$ limit of the result (\ref{kappa_exact}) below in a direct $1/N_f$ expansion, with $k$ taken to be of order $N_f$. Note that the first term in Eq.~\eqref{kappa_exact} is of order $N_f^1$, while the second term is of order $N_f^0$.

Our $1/N_f$ expansion also yields a simple interpretation of the two terms in \eqref{kappa_exact}. The leading term of order $N_f$ is the contribution of free Dirac fermions, where we assume that the Chern-Simons term in $A_\mu$ was generated by integrating out massive Dirac fermions. The contribution to the universal scaling function $\mathcal{K}$ by the free Dirac fermions is specified by (\ref{eq:kxy_gt}) and plotted in Fig.~\ref{fig:Kxy}.
The subleading term of order $N_f^0$ is the contribution of the fluctuations of $A_\mu$. We will exploit this interpretation when we consider the doped case.

We also consider the quantum critical limit, $|m|/T \rightarrow 0$, when neither an exact computation of $\kappa_{xy}$ is possible, and nor is $\kappa_{xy} (\hbar/(\pi k_B^2))$ expected to be quantized at a rational value. In this case, we obtain
\beq
\kappa_{xy} = \frac{\pi k_B^2 T}{6 \hbar} \left[ 2 k + \mathcal{O}(N_f^0)   \right] \quad, \quad \frac{|m|}{T} \rightarrow 0\,.
\label{kappa_qc}
\eeq
The computation of the $\mathcal{O}(N_f^0)$ number requires a lengthy numerical computation which we will outline, but not carry out to completion. We note that we do not expect $\kappa_{xy}$ in the limit $|m|/T \rightarrow 0$ to be related to any gravitational anomaly
or contact terms \cite{Closset:2012vg,Closset:2012vp}; the latter are evaluated at $T=0$, and not in the limit required for a quantum critical transport coefficient, with frequencies much smaller than $T$ \cite{KDSS97,SS97}.

\subsection{Pseudogap at nonzero doping}
\label{sec:pseudogap}

We will describe the pseudogap by essentially the same theory as that used in \refcite{SCWFGS}, which was successfully compared with numerical studies of the Hubbard model \cite{SCWFGS,WSCSGF,Scheurer_tri} and photoemission experiments on an electron-doped cuprate \cite{He2019}. In the limit of the insulating state, and in the vicinity of the onset of semion topological order in the presence of N\'eel order as discussed in Section~\ref{sec:undoped}, this theory can be related  \cite{THNP} to one of the theories which are equivalent to (\ref{eq:L_CS}) after duality---a SU(2) gauge theory at Chern-Simons level 1, coupled to a complex scalar which is a SU(2) fundamental. While the fermionic SU(2) theory at level $-1/2$ in (\ref{eq:L_CS}) was useful in describing $\kappa_{xy}$ in the insulator \cite{THNP}, the complex scalar SU(2) theory is far more convenient in the doped case. This is because the latter theory has fermionic charge carriers, and this allows easy access to a metallic state at nonzero doping.

The pseudogap metal is described by transforming to a rotating reference frame in spin space \cite{SS09}, which results in a SU(2) gauge theory. The fluctuating spin density wave order acts like a Higgs field, which breaks the SU(2) invariance down to U(1). Coupled to the U(1) gauge field, $a_\mu$, we have bosonic spinons and fermionic chargons $f_p$ with U(1) gauge charges $p=\pm 1$. We focus on the fermionic chargons, as they form Fermi pockets with charged gapless excitations on the Fermi surface.
We write down a simple effective theory for these chargons \cite{Lee89,SCWFGS,WSCSGF}:
\beq
\begin{split}
\mathcal{L}^{}_f = &\sum_{v=1,2} \sum_{p=\pm 1} f_{pv}^\dagger \left( \frac{\partial}{\partial \tau} - \mu - i p a^{}_\tau \right.\\
&-\left.\frac{ (\vect{\nabla} - i p \vect{a} - i e \vect{A}_{\rm em} )^2}{2 m^\ast}+v_{\rm dis}(\vect{r})  \right ) f^\pdagger_{pv}
\label{eq:Lf}
\end{split}
\eeq
Here, $v$ is a valley index, $m^\ast$ is the effective mass of the fermions (we have ignored mass anisotropies), $\mu$ is a chemical potential, $a_\mu$\,$=$\,$(a_\tau, \vec{a})$ is the emergent U(1) gauge field, and $\vect{A}_{\rm em}$ is the fixed background electromagnetic gauge field associated with the applied magnetic field $B = \hat{\boldsymbol{z}} \cdot (\vect{\nabla} \times \vect{A}_{\rm em})$. We have included a disorder potential $v_{\rm dis} (\vect{r})$, because we will consider Hall transport in the weak-field regime $\omega_c \tau \ll 1$, where $\omega_c$ is the cyclotron frequency and $\tau$ is the elastic scattering time associated with the disorder.

First, let us ignore the internal gauge field $a_\mu$. Then, the $f$ fermions form a conventional Fermi liquid, and for $\omega_c \tau \ll 1$, the electrical and thermal Hall responses are given by familiar expressions involving the Wiedemann-Franz relation
\beq
\rho^{}_{xy} = \frac{B}{nec}, \quad \sigma_{xy} \approx \frac{\rho^{}_{xy}}{\rho_{xx}^2} , \quad \kappa_{xy}^0 =  \frac{\pi^2 T}{3} \left( \frac{k^{}_B}{e} \right)^2 \sigma^{}_{xy} \,, \label{WF}
\eeq
where $n$ is the total density of the $f$ fermions.
Now, let us consider the contribution of $a_\mu$ to the thermal Hall response. We will compute this by a simple Maxwell-Chern-Simons action for $a_\mu$
\beq
\mathcal{L}^{}_a = \frac{K^{}_1  }{2} ( \vect{\nabla} \times \vect{a} )^2 + \frac{K^{}_2  }{2} (\vect{\nabla} a_\tau - \partial_\tau \vect{a})^2 - \frac{i \sigma^{}_{xy}  }{2e^2} \epsilon^{}_{\mu\nu\lambda}a^{}_\mu \partial^{}_\nu a^{}_\lambda. \label{MCS}
\eeq
We assume that the predominant contribution to the Maxwell terms arises from integrating out the gapped spinons. Integrating out the fermionic chargons introduces the Chern-Simons term in (\ref{MCS}), proportional to the Hall conductivity of the fermions in (\ref{WF}); such a term is also permitted under the symmetry constraints on this gauge theory of doped antiferromagnets \cite{Balents:2006wm,scheurer2018orbital}. In general, because of the presence of disorder,
the couplings $K_{1,2}$ will also be functions of spatial position; we replace them by their spatial average, and do not expect fluctuations to significantly modify the results presented here.
The fermions also introduce singular terms in the transverse gauge field propagator arising from Landau damping \cite{HLR}, so that a more complete effective action is
\beq
\mathcal{S}^{}_a = \int d^2 x\, d \tau \, \mathcal{L}_a + \int
\frac{d^2k\, d\omega}{8 \pi^3} \, \gamma^{}_k |\omega| \, [\vect{a}^T (k, \omega)]^2\,, \label{Landaudamping}
\eeq
where $\gamma_k \sim 1/k $ for $k v_F \tau \gg 1$, and $\gamma_k \sim \mbox{constant}$ for $k v_F \tau \ll 1$.
Although the term in (\ref{Landaudamping}) could make a significant contribution to the thermal Hall effect, we leave an analysis of its effects to future work.

Computing the thermal Hall response of the Maxwell-Chern-Simons theory $\mathcal{L}_a$ in Section~\ref{sec:mcs}, we find that it yields a correction $\kappa_{xy}^1$, which has the {\it opposite} sign from $\kappa_{xy}^0$ in (\ref{WF}). This sign change is similar to that in (\ref{kappa_exact}) between the $\mathcal{O}(N_f)$ term (from the fermions) and the $\mathcal{O}(N_f^0)$ term (from the gauge field). The universal function $\mathcal{K}$ in (\ref{kappauniv}) for the Maxwell-Chern-Simons theory is specified in (\ref{eq:kappaxy_gauge}) as a function of the `topological mass' $m_t =  \sigma_{xy}/(e^2 K_2) $, and is plotted in Fig.~\ref{fig:kappaxy2}. Note that the universal function \eqref{eq:kxy_gt} for Dirac fermions in Fig.~\ref{fig:Kxy} does not reduce to the gauge-field function in (\ref{eq:kappaxy_gauge}) and Fig.~\ref{fig:kappaxy2} by a rescaling of axes: this is evidence that the $T>0$ thermal Hall conductivity is a bulk property, and is not specified by any topological field theory or gravitational anomaly.

We begin our analysis by describing the thermal Hall response of two free theories: a free Dirac fermion in Section~\ref{sec:dirac}, and free Maxwell-Chern-Simons theory in Section~\ref{sec:mcs}. The results of the Maxwell-Chern-Simons theory apply directly to the effective theory for the doped pseudogap phase in Eq.~(\ref{MCS}).
We will combine Sections~\ref{sec:dirac} and \ref{sec:mcs} to obtain results for the Dirac Chern-Simons theory (\ref{eq:L_CS}) in the $1/N_f$ expansion in Section~\ref{sec:dcs}.

\section{Free Dirac fermion}
\label{sec:dirac}

In order to obtain a finite thermal Hall effect, time-reversal symmetry must be broken: this can be achieved by either an external magnetic field or intrinsic magnetic ordering \cite{katsura2010theory}. However, initial attempts to calculate this response based on direct application of the Kubo formula were found to suffer from unphysical divergences at zero temperature \cite{agarwalla2011phonon, matsumoto2011theoretical}.
This is because in a system breaking time-reversal symmetry, a temperature gradient drives not only the transport (heat) current, but also an experimentally unobservable circulating current \cite{smrcka1977transport,cooper1997thermoelectric}. Both contributions are present in the microscopic current density calculated by the standard linear response theory, necessitating a proper subtraction of the circulating component. Ref.~\onlinecite{qin2011energy} showed that the electromagnetic and gravitomagnetic energy magnetizations \cite{luttinger1964theory, ryu2012electromagnetic} naturally emerge as corrections to the thermal transport coefficients, removing the aforementioned divergences in the process. A subtlety pointed out in Ref.~\onlinecite{Kapustin19} is that the energy magnetization and the thermal Hall coefficient are relative: only the difference between two systems are physically meaningful. We choose to normalize $\kappa_{xy}$ such that the $\kappa_{xy}/T\to 0$ as $m/T\to 0$, i.e., the vacuum has zero thermal Hall coefficient.

We now present details of the computation of the thermal Hall coefficient of a free Dirac fermion with a mass $m$ which can be scanned through zero at $T>0$. This is the theory (\ref{eq:L_CS}) without the gauge field $A_\mu$. While we consider a single two-component Dirac fermion, note that, because of the SU(2) gauge index, the theory (\ref{eq:L_CS}) has $2N_f$ such fermions. In the following, we determine both the Kubo part and the magnetization separately, identifying precisely what the transport currents and the magnetizations are, and illustrating how they can be evaluated for a general continuum theory.

\subsection{Transport contribution from the Kubo formula}

\begin{figure}[htb]
\includegraphics[width=\linewidth]{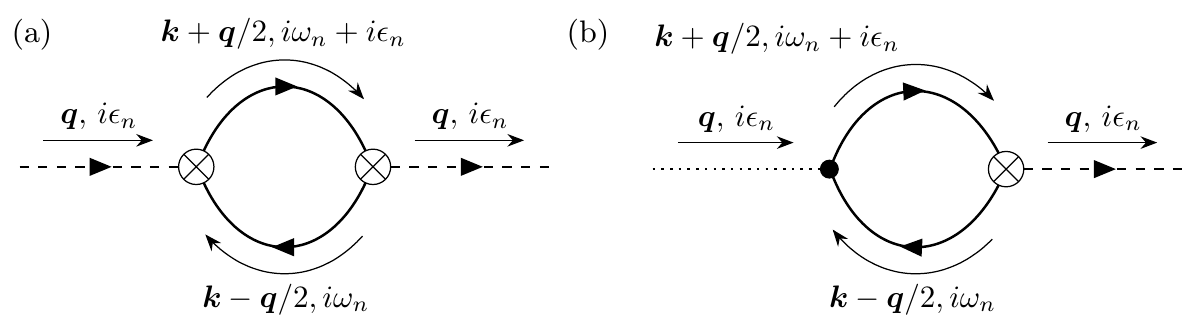}
\caption{\label{fig:mag}\label{fig:fbubble}The fermion polarization bubbles that give the mean-field Kubo (left) and internal magnetization (right) contributions to the thermal Hall conductivity; the crossed circles represent thermal current vertices.}
\end{figure}

The leading contribution to $\kappa_{xy}$ is given by a single fermion polarization bubble shown in Fig.~\ref{fig:fbubble}.

Summing over the internal momentum $\vect{k}$ and Matsubara frequency $i \omega_n$, this diagram evaluates to
\begin{alignat}{1}
&\Pi^\textsc{q}_{xy} (\vect{q}, i \epsilon_n) = \frac{1}{\beta \,V}\sum_{\vect{k}, i\omega_n} \mathrm{Tr} \Big[ \mathcal{V}^\textsc{q}_x (-\vect{q}, -i \epsilon_n)\nonumber\\
&\mathcal{G} (\vect{k}-\vect{q}/2, i \omega_n)
\mathcal{V}^\textsc{q}_y (\vect{q}, \,i \epsilon_n)\, \mathcal{G} (\vect{k}+\vect{q}/2, i \omega_n + i \epsilon_n)\Big],
\end{alignat}
where
$G^{\pdagger}_\Psi (\vect{k}, i\omega_n) =
1/(-i\, \omega_n + \vect{\sigma} \cdot \vect{k} + m\, \sigma^z)$ is the free fermion Green's function, and $\mathcal{V}^\textsc{q}_a (\vect{q}, i \epsilon_n) = \sigma^a\, (i \omega_n + i \epsilon_n/2)$ is the heat/energy-current vertex, derived in Appendix~\ref{app:feynmanrules}. The response function is defined as
\begin{alignat*}{1}
L^{}_{xy} &= \frac{1}{\epsilon_n} \Pi^\textsc{q}_{xy} (\vect{q}, i \epsilon_n)\\
&= \frac{T}{2\epsilon_n\,V} \sum_{\vect{k}, i\omega_n} \frac{- (\epsilon_n+2 \omega_n )^2 }{ k_x^2+k_y^2+m^2+\omega_n ^2} \\
&\times\frac{2 k_x k_y + m\, \epsilon_n}{k_x^2+k_y^2+m^2+(\epsilon_n+\omega_n )^2}
\mbox { as } \vect{q} \rightarrow 0,
\end{alignat*}
specializing to the case of zero external momentum.
The numerator of the polarization tensor $\Pi^\textsc{q}_{\mu \nu}$ consists of a part proportional to $k_\mu k_\nu$  and a term $ \sim \delta_{\mu \nu}$; we can drop the former because it is odd in $k_x$ and $k_y$ and hence, vanishes upon integration over all momenta. Performing the Matsubara summation and converting the momentum sum to an integral, we get
\begin{alignat}{1}
L^{}_{xy} &=-\frac{1}{\epsilon^{}_n} \int \frac{\mathrm{d}^2 \vect{k}}{(2\pi)^2} \,{\displaystyle\frac{m \,\epsilon_n}{2\, \xi_{\vect{k}} }\tanh \left({\displaystyle \frac{\beta  \xi_{\vect{k}} }{2}}\right) },
\end{alignat}
where $\xi_{\vect{k}} = \sqrt{k_x^2+k_y^2+m^2}$. Finally, introducing the shorthand $u = \beta\, \xi_{\vect{k}}$, we have
\begin{alignat}{1}
\nonumber L^{}_{xy}& = -\int_{\beta \lvert m \rvert}^\infty \frac{\mathrm{d} u}{4\pi\,\beta} m\,\tanh\left(\frac{u}{2} \right)\\
&= -\frac{m}{2 \pi\,\beta} \ln \left[\cosh \left(\frac{u}{2}\right)\right] \bigg \vert_{\beta \lvert m \rvert}^{\infty}.
\end{alignat}
A few comments are in order about this result. First, we have regulated the integral by introducing a UV cutoff $\Lambda$ but, as we shall see, this drops out eventually. Further, to obtain the DC response, we need to analytically continue to real frequencies $i \epsilon_n \rightarrow \epsilon + i 0^+$, and then take the limit $\epsilon \rightarrow 0$ \textit{after} $\vect{q} \rightarrow 0$. The thermal Hall coefficient is then \cite{qin2011energy, matsumoto2014thermal}
\begin{alignat}{1}
\kappa^{\mathrm{Kubo}}_{xy} &\equiv  \frac{L^{}_{xy}}{T} \\
\nonumber&=\frac{m}{2 \pi}\left\{\ln \left[\cosh \left(\frac{\beta \,\lvert m \rvert}{2}\right)\right] - \ln \left[\cosh \left(\frac{\beta\,\Lambda}{2}\right)\right]\right\}.
\end{alignat}

\subsection{Internal magnetization}

The second contribution to the conductivity comes from the circulating heat current. The zero-field heat magnetization can be calculated from the differential equation \cite{qin2011energy}
\begin{equation}
\label{eq:rel}
2 \vect{M}_Q^\pdagger - T \frac{\partial \vect{M}_Q^\pdagger}{\partial T} = \frac{1}{2 i} \vect{\nabla}_{\vect{q}}\times \left \langle \hat{K}^\pdagger_{-\vect{q}}; \hat{J}^Q_{\vect{q}}  \right \rangle_0 \bigg\vert_{\vect{q} \rightarrow 0},
\end{equation}
where $\hat{K}_{\vect{q}}$ is the Fourier transform of $\hat{K} ({\vect{r}}) = h ( {\vect{r}} ) - \mu\, \hat{n} ( {\vect{r}} )$, $h$ and $\hat{n}$ being the local energy and number densities, respectively. The correlator on the RHS is evaluated in the static limit, i.e. $\vect{q}\rightarrow 0$ \textit{after} $\epsilon\rightarrow 0$. We also point out to readers that Eq.~\eqref{eq:rel}, cited from Ref.~\onlinecite{qin2011energy}, only applies to systems whose energy current depends on the gravitational field in a particular way, while the general formalism is discussed in Ref.~\onlinecite{Kapustin19}.

Equipped with the structure of this modified vertex from Appendix~\ref{app:feynmanrules}, we now evaluate Eq.~\eqref{eq:rel} piece by piece. Consider the first term in the curl; retaining only the terms even in the internal momentum \footnote{This is allowed since we focus on the limit $\vect{q}=0$. More formally, this corresponds to expanding the full integrand in powers of $q_x$ and $q_y$ and then keeping only those terms of $\mathcal{O}(q_x)$ that would survive the $\vect{k}$ integration.}, we get
\begin{alignat}{1}
&\nonumber\partial_{q_x} \left \langle \hat{K}^\pdagger_{-\vect{q}}; \hat{J}^Q_{y, \vect{q}}  \right \rangle \\ \nonumber
&= \partial_{q_x} \Bigg(\frac{1}{\beta V} \sum_{\vect{k}, i \omega_n} \frac{ 2i\, m\, q_x (\epsilon_n+2 \omega_n )^2 }{4\,\left(\lvert \vect{k} - \vect{q}/2 \rvert^2+ m^2+\omega_n^2 \right) }\\ \nonumber
&\qquad\,\,\,\,\times\frac{1}{\left(\lvert \vect{k} + \vect{q}/2 \rvert^2+m^2+(\epsilon_n+\omega_n )^2\right)}\Bigg)\Bigg\vert_{\epsilon_n=0}\\
\nonumber &=  i \int \frac{\mathrm{d}^2 \vect{k}}{(2\pi)^2} \frac{m \left(\beta\,  \xi_{\vect{k}}+\sinh \left(\beta\,  \xi_{\vect{k}}\right)\right) \text{sech}^2\left(\beta\,  \xi_{\vect{k}}/2\right)}{4\, \xi_{\vect{k}}}\\
& = i \int_{\beta \lvert m \rvert}^\infty \frac{\mathrm{d}u}{8 \pi\,\beta}\,m\, u\, (u+\sinh (u))\, \text{sech}^2\left(u/2\right) \nonumber\\
&= \frac{i m \,u }{4 \pi\,\beta } \tanh \left(\frac{u}{2}\right) \bigg\vert_{\beta \lvert m \rvert}^{\infty}.
\label{eq:comp1}
\end{alignat}
In the second line, we evaluate the sum at $\epsilon_n=0$ and then take the $q_x$ derivative.
Similarly, as expected by symmetry,
\begin{alignat}{1}
\label{eq:comp2}
\partial_{q_y} \left \langle \hat{K}^\pdagger_{-\vect{q}}; \hat{J}^Q_{x, \vect{q}}  \right \rangle &= - \frac{i\, m \,u }{4 \pi\,\beta} \tanh \left(\frac{u}{2}\right) \bigg\vert_{\beta \lvert m \rvert}^{\infty}.
\end{alignat}
Plugging Eqs.~\eqref{eq:comp1} and \eqref{eq:comp2} back into Eq.~\eqref{eq:rel}, we have
\begin{align}
&2 \vect{M}_Q^\pdagger - T \frac{\partial \vect{M}_Q^\pdagger}{\partial T} = \nonumber\\
&\frac{m}{4 \pi} \left [\Lambda \tanh \left( \frac{\beta\,\Lambda}{2} \right) - \lvert m \rvert \tanh \left(\frac{\beta \lvert m \rvert}{2}\right) \right],
\end{align}
which can be solved for $\vect{M}_Q$ to obtain
\begin{align}
\vect{M}_Q^\pdagger &= c^\pdagger_1 T^2 - \frac{m T}{4 \pi } \Bigg[\frac{2 \text{Li}_2\left(-e^{-\Lambda /T}\right)}{\Lambda/T }-\frac{2 \text{Li}_2\left(-e^{-\lvert m \rvert/T}\right)}{\lvert m \rvert /T}\nonumber\\
&+\frac{\lvert  m \rvert - \Lambda }{2 T}+2\ln \left({\displaystyle\frac{e^{-\lvert m \rvert /T}+1}{e^{-\Lambda/T}+1}}\right)\Bigg],
\end{align}
where $c^\pdagger_1$ is an arbitrary constant. Now, $\kappa^{\mathrm{tr}}_{xy} = \kappa^{\mathrm{Kubo}}_{xy} + 2 \vect{M}_Q / T$. Collecting the terms proportional to $\Lambda$, we get
\begin{align}
\lim_{\Lambda \rightarrow \infty} \frac{m}{4 \pi} \Bigg[&-2 \ln \left( \cosh \left(\frac{\beta \Lambda}{2}\right)\right) + \beta \Lambda  \nonumber\\
&- \frac{4 \text{Li}_2\left(-e^{-\beta\,\Lambda }\right)}{\beta \,\Lambda}+ 4 \ln \left(e^{-\beta\,\Lambda}+1 \right)\Bigg] = \frac{m}{2\pi} \ln 2,
\end{align}
the first two terms cancel out the UV divergences and all dependencies on the cutoff $\Lambda$ drop out. Thus, the physical thermal Hall conductivity is given by
\begin{equation}
\label{eq:kxy_gt}
\kappa^{\mathrm{tr}}_{xy} = 2 c^\pdagger_1 T+\frac{m}{2 \pi } \left[\frac{2 \text{Li}_2\left(-e^{-\lvert m \rvert/T}\right)}{\lvert m \rvert /T} -  \ln \left(e^{-\lvert m \rvert /T}+1 \right) \right] ,
\end{equation}
where the first part comes from the magnetization and the last piece is the Kubo contribution.

\begin{figure}
\includegraphics[width=\linewidth]{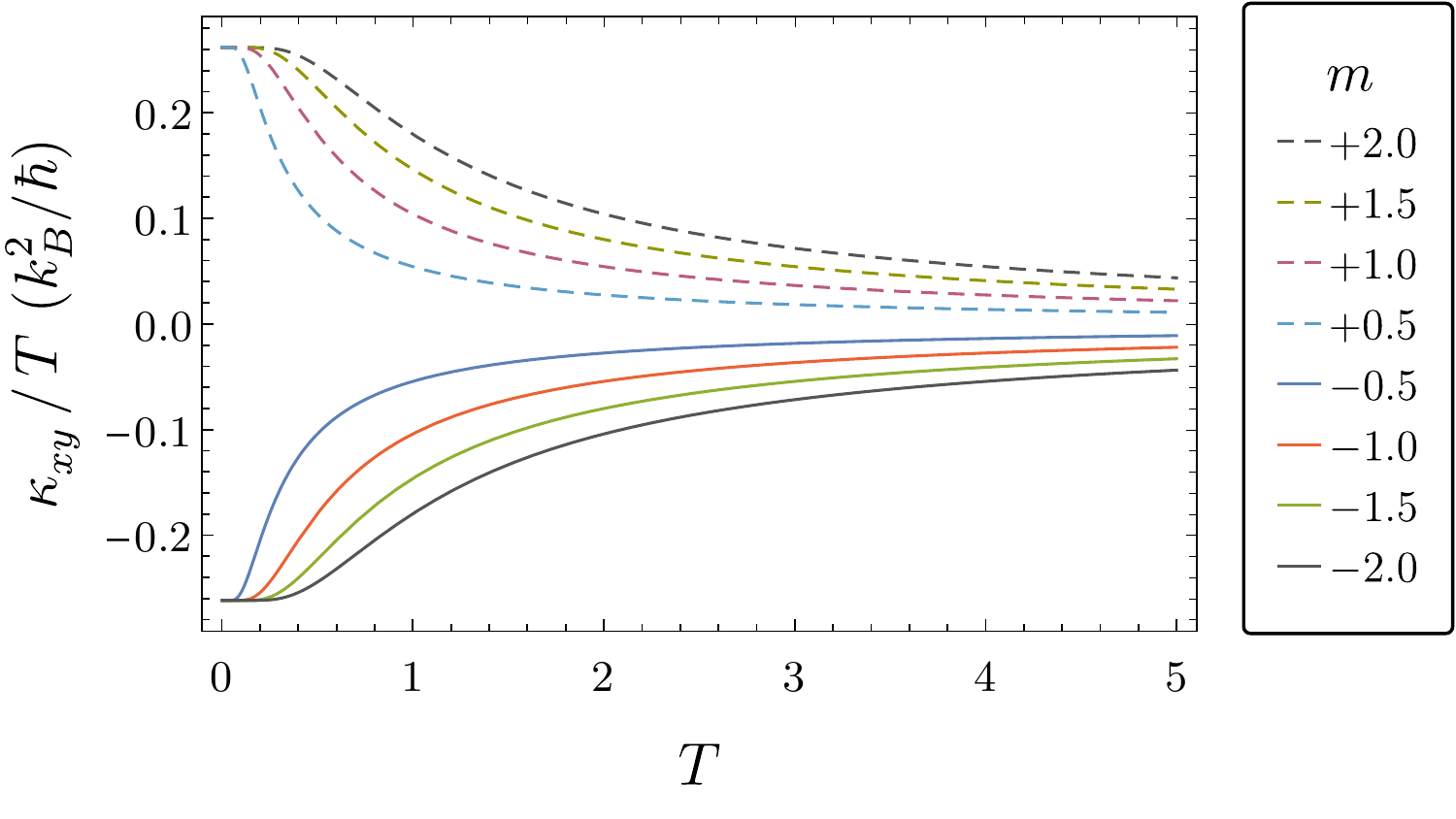}
\caption{The thermal Hall conductivity of a free Dirac fermion of mass $m$. The values as $T \rightarrow 0$ are $\pm \pi/12$.}
\label{fig:Kxy}
\end{figure}

At this point, the constant $c^\pdagger_1$ arising from the solution of the differential equation can be determined as follows. We have seen above that $\kappa_{xy}/T$ is a function of the dimensionless variable $\lvert m \rvert /T$ alone. Therefore, taking the limit $m \rightarrow 0$ or equivalently, $T \rightarrow \infty$  (where we know \textit{a priori} that $\kappa_{xy}/T$ should go to zero), Eq.~\eqref{eq:kxy_gt} reduces to
\begin{equation}
\kappa^{\mathrm{tr}}_{xy} = 2 c^\pdagger_1 T +  \frac{\sign(m) }{2\pi} \left(-2 \cdot \frac{\pi^2}{12} \right) T \equiv 0
\end{equation}
and the last condition implies that $c^\pdagger_1 = \sign(m)\, \pi/24$. As a result, when $T \rightarrow 0$ keeping $m \ne 0$ and fixed, we obtain $\kappa^{\mathrm{tr}}_{xy}/T = \sign(m)\, \pi/12$. The dependence of $\kappa^{\mathrm{tr}}_{xy}$ on temperature and mass is shown in Fig.~\ref{fig:Kxy}.

\section{Maxwell-Chern-Simons Theory}
\label{sec:mcs}
In this section, we consider the framing anomaly and thermal Hall response of the U(1) Maxwell-Chern-Simons theory (\ref{MCS}). Our discussion will be restricted to the level of an effective theory and we do not attempt to extract the microscopic values of $K_1,K_2$. As we will see, the effective theory already provides a satisfactory interpolation between the two topological phases.

The Maxwell-Chern-Simons theory has a speed of `light' $c_0=\sqrt{K_1/K_2}$. Since there is no other velocity scale in the theory, we can set $c_0=1$ (see also Appendix~\ref{app:wrong}). The MCS theory \eqref{MCS} takes the following relativistic form in real time:
\begin{equation}
    \mathcal{L}=\frac{k}{4\pi}\varepsilon^{\mu\rho\nu}a_\mu\partial_\rho a_\nu-\frac{1}{4g}f_{\mu\nu}f^{\mu\nu},
\end{equation}
where $f_{\mu\nu}=\partial_\mu a_\nu-\partial_\nu a_\mu$. Here, the coupling $g=1/K_2$ has dimensions of energy in 2+1D, and $k=2\pi\sigma_{xy}/e^2$ is the Chern-Simons Level. The large-$N_f$ limit of fermion flavor implies $k=\mathcal{O}(N_f)$ and $g=\mathcal{O}(1/N_f)$. If we are interested in the thermal Hall effect of the gapped phases, we should take $g$ to be the largest energy scale and send $g\to\infty$.

Chern-Simons theory has a gravitational anomaly called the framing anomaly. It is well known that Chern-Simons theory is topological and does not couple to spacetime geometry on the classical level. \citet{Witten} pointed out that at the quantum level, the theory inevitably couples to a metric because of the gauge-fixing procedure. However, this is not adequate for writing down a sensible stress tensor, because any vertex function due to the gauge-fixing procedure is longitudinal and thus, vanishes when contracted with the physical transverse propagator. Here, we will try an alternative method, by considering the Maxwell-Chern-Simons theory. The Maxwell term serves as a UV regulator and it enables us to write down a stress-tensor vertex. According to \citet{Witten}, the gravitational anomaly appears as a gravitational Chern-Simons term
\begin{equation}\label{eq:CSg}
  CS_g[g_{\mu\nu}]=\frac{c}{96\pi}\int \rm{tr}\left(\Gamma\wedge\rd\Gamma+\frac{2}{3}\Gamma\wedge\Gamma\wedge\Gamma\right),
\end{equation} where $\Gamma$ is the Christoffel symbol associated with the metric $g_{\mu\nu}$.
 The prefactor $c$ is the (chiral) central charge
\begin{equation}c=-\frac{{\rm{dim}}(G)\,k}{(|k|+c_2(G))},\label{eq:centralcharge}\end{equation}
where $c_2(G)$ is the dual Coxeter number of the gauge group $G$.

 In what follows, we perturbatively compute the gravitational anomaly and thermal Hall effect of the above MCS theory in the large-$k$ ($k\propto N_f$) limit. Following \cite{Asorey}, we calculate the stress tensor-stress tensor correlation function $\Pi^{\mu\nu;\rho\lambda}(x,t)=-i\langle T^{\mu\nu}(x,t) T^{\rho\lambda}(0,0)\rangle$ of the MCS theory. We can interpret $\Pi^{\mu\nu;\rho\lambda}$ as the effective action (up to a minus sign) $S_\text{eff}[g_{\mu\nu}]$ of metric $g_{\mu\nu}$ in a weakly curved background $g_{\mu\nu}=\eta_{\mu\nu}+h_{\mu\nu},~|h_{\mu\nu}|\ll|\eta_{\mu\nu}|$. 
 We will show that at zero temperature $\Pi^{\mu\nu;\rho\lambda}$ agrees with the gravitational Chern-Simons term \eqref{eq:CSg}.


This gravitational anomaly is proportional to the thermal Hall coefficient at the next-leading large-$N_f$ order via the relation
\begin{equation}\label{}
  \kappa^{}_{xy}=\frac{\pi}{6}c\,T.
\end{equation}
It is argued in \refcite{Stone12} that a gravitational Chern-Simons term cannot give rise to a thermal Hall effect from the Kubo formula because it contains three derivatives rather than one. In our calculation, we find that the thermal Hall effect actually arises from the finite-temperature part of $\Pi^{\mu\nu;\rho\lambda}$, which comes from the same diagrams as the gravitational anomaly and contains only one derivative.

At the next-leading large-$N_f$ order, our approach works both for Abelian and non-Abelian theories because $c={\rm dim}(G)+\mathcal{O}(1/N_f)$, and we simply include ${\rm dim}(G)$ copies of gauge fields, which are noninteracting at this order.


\subsection{Propagator}
    We add a gauge fixing term $\mathcal{L}_{gf}=(\partial_\mu a^\mu)^2/(2\xi g)$ and work out the propagator.
Some algebra leads to
\begin{align}
    &S=\nonumber\int\frac{\rd^3 p}{(2\pi)^3}\frac{1}{2g}\\
   \nonumber &a_\mu(-p)\left(p^\mu p^\nu-\eta^{\mu\nu}p^2+\frac{p^\mu p^\nu}{\xi}+\frac{k g}{2\pi}\varepsilon^{\mu\nu\rho}ip_\rho\right)a_\nu(p).
\end{align}
    The propagator is thus
\begin{equation}
    D^{\mu\nu}(p)=\frac{-g}{p^2-m_t^2}\left[P^{\mu\nu}(p)+\frac{m_t}{p^2}\varepsilon^{\mu\nu\rho}ip_\rho\right]+\frac{\xi}{(p^2)^2}p^\mu p^\nu,
\end{equation} where the topological mass $m_t$\,$=$\,$kg/(2\pi)$, and $P^{\mu\nu}(p)=\eta^{\mu\nu}-p^\mu p^\nu/p^2$.

\subsection{Stress-tensor vertex}

\begin{figure}
{\centering
\includegraphics[width=0.8\linewidth]{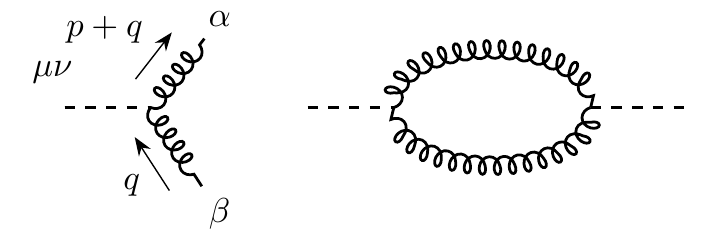}}
\caption{\label{fig:stress_tensor_vertex}Left: Vertex diagram of the stress tensor. Right: Diagram of the stress tensor-stress tensor correlation function.}
\end{figure}

    In this section, we work out the stress-tensor vertex as shown in Fig.~\ref{fig:stress_tensor_vertex}.
    The stress tensor is given by the Maxwell term, which is
\begin{equation}\label{eq:stress_tensor}
    T^{\mu\nu}=\frac{-1}{g}\left[f^{\mu\rho}f^{\nu}_{~\rho}-\frac{1}{4}\eta^{\mu\nu}f_{\alpha\beta}f^{\alpha\beta}\right].
\end{equation}

   It is worth noticing that the stress tensor above can be derived solely from translation symmetry and gauge invariance, without reference to Lorentz symmetry (see Appendix.~\ref{app:feynmanrules}).

    We write the vertex function of Fig.~\ref{fig:stress_tensor_vertex} as
\begin{equation}-\Gamma_{\mu\nu;\alpha\beta}(p+q,q)/g,
\end{equation} where
\begin{widetext}
\begin{equation}\label{eq:Gamma}
\begin{split}
    \Gamma^{\mu\nu;\alpha\beta}(p+q,q)&=\left[(p+q)^\mu q^\nu+(p+q)^\nu q^\mu-\eta^{\mu\nu}(p+q)\cdot q\right]\eta^{\alpha\beta} +(p+q)\cdot q\left(\eta^{\mu\alpha}\eta^{\nu\beta}+\eta^{\mu\beta}\eta^{\nu\alpha}\right)\\
   &-\Big[q^\alpha\left(\eta^{\beta\mu}(p+q)^\nu+\eta^{\beta\nu}(p+q)^\mu\right) +(p+q)^\beta\left(\eta^{\alpha\mu}q^\nu+\eta^{\alpha\nu}q^\mu\right)-\eta^{\mu\nu}q^\alpha(p+q)^\beta\Big].
\end{split}
\end{equation}
\end{widetext}
\subsection{Stress tensor-stress tensor correlation function}

We compute the following stress tensor-stress tensor polarization function:

\begin{equation}\label{eq:Pi_def}
    \Pi^{\mu\nu;\rho\lambda}(x,t)=-i\langle T^{\mu\nu}(x,t) T^{\rho\lambda}(0,0)\rangle.
\end{equation}
  In general, the full polarization function should also contain contact terms such as $\langle\delta T^{\mu\nu}(x,t)/\delta h_{\rho\lambda}(0,0)\rangle$, but those terms are symmetric in $\mu\rho$ and independent of external momentum, so they do not contribute to either the gravitational anomaly or the thermal Hall effect.
We only need to consider the single bubble diagram in Fig.~\ref{fig:stress_tensor_vertex}, which yields
\begin{equation}\label{eq:Pi_TT}
\begin{split}
 &   \Pi^{\mu\nu;\rho\lambda}(p)=\int\frac{\rd^3 q}{(2\pi)^3}\frac{-i}{2g^2}\Gamma^{\rho\lambda;\alpha\beta}(p+q,q)iD_{\alpha'\alpha}(p+q)\\ &\times\Gamma^{\mu\nu;\beta'\alpha'}(q,p+q)iD_{\beta\beta'}(q).
\end{split}
\end{equation}
    Here, we have included a symmetry factor of $1/2$.

We want to extract the part which is antisymmetric in $\mu\rho$ and symmetric in $\nu\lambda$, which should ultimately lead to a gravitational Chern-Simons term and thermal Hall effect:
\begin{equation}\label{eq:Pi_AS}
\Pi_\text{AS}^{\mu\nu;\rho\lambda}=\frac{1}{4}\left(\Pi^{\mu\nu;\rho\lambda}-\Pi^{\rho\nu;\mu\lambda}+\Pi^{\mu\lambda;\rho\nu}-\Pi^{\rho\lambda;\mu\nu}\right).
\end{equation}

\subsection{$T=0$: Gravitational Chern-Simons term}

At zero temperature, the integrand has Lorentz symmetry, and we can evaluate $\Pi_\text{AS}$ using Feynman parameters:
\begin{equation}
\begin{split}
    &\frac{1}{q^2(p+q)^2(q^2-m_t^2)((p+q)^2-m_t^2)}=\\
    &6\int_0^1\rd x_1\rd x_2\rd x_3\rd x_4\, \delta\left(\sum x-1\right)\frac{1}{(l^2-\Delta)^4},
\end{split}
\end{equation}
where
\begin{eqnarray}
    l&=&q+(x_1+x_3)p,\\
    \Delta&=&(x_1+x_2)m_t^2-(x_1+x_3)(1-x_1-x_3)p^2.
\end{eqnarray}

After some algebra with Mathematica, we have
\begin{equation}
\begin{split}
    &\Pi_\text{AS}^{\mu\nu;\rho\lambda}(p)=-\varepsilon^{\mu\rho\sigma}p_\sigma(\eta^{\nu\lambda}p^2-p^\nu p^\lambda)\\
    &\times6\int_0^1\rd x_1\rd x_2\rd x_3\rd x_4\, \delta\left(\sum x-1\right)\left(\frac{-m_t}{15}\right)\\
&\times\int \frac{\rd^3 l}{(2\pi)^3}\frac{l^4[10-63(x_1+x_3)(1-x_1-x_3)]+\mathcal{O}(l^2)}{(l^2-\Delta)^4}.
\end{split}
\end{equation}
To obtain the gravitational Chern-Simons term, we isolate the topological contributions by taking the limit $m_t\to\infty$. Note that only the $l^4$ term written above can survive the $m_t\to\infty$ limit. The integral can be evaluated using dimensional regularization. The result is
\begin{equation}\label{eq:Pi_ASres}
\Pi_\text{AS}^{\mu\nu;\rho\lambda}(p)=\frac{-i}{48\pi}\sgn (m_t)\varepsilon^{\mu\rho\sigma}p_\sigma(\eta^{\nu\lambda}p^2-p^\nu p^\lambda),
\end{equation}
employing the integral formula
\begin{equation}
\int \frac{\rd^d l_E}{(2\pi)^d}\frac{(l_E^2)^a}{(l_E^2+D)^b}=\frac{\Gamma(b-a-\frac{d}{2})\Gamma(a+\frac{d}{2})}{(4\pi)^{d/2}\Gamma(b)\Gamma(\frac{d}{2})}\frac{1}{D^{b-a-\frac{d}{2}}}.
\end{equation}
We can compare the above result to the gravitational Chern-Simons term \eqref{eq:CSg},
which, to quadratic order in $h$, reduces to
\begin{equation}\label{eq:CSgh}
\begin{split}
    CS_g[h]=-\frac{c}{192\pi}&\int\frac{\rd^3p}{(2\pi)^3}h_{\mu\nu}(-p)\varepsilon^{\mu\rho\sigma}(ip_\sigma)\\
    &(p^2\eta^{\nu\lambda}-p^\nu p^\lambda)h_{\rho\lambda}(p).
\end{split}
\end{equation}
The correlation function is related to $CS_g$ by
\begin{equation}\label{eq:TT=ddCSg}
\begin{split}
   \Pi_\text{AS}^{\mu\nu;\rho\lambda}(p)&= -i\langle T^{\mu\nu}(p)T^{\rho\lambda}(-p)\rangle_{AS}\\
   &=-\frac{\delta^2 CS_g[h]}{\delta h_{\mu\nu}(-p)\,\delta h_{\rho\lambda}(p)}+(\text{symmetrization}).
\end{split}
\end{equation}
    When evaluating the variation, we get a trivial factor of $2$ because $CS_g$ is quadratic in $h$. There is another hidden factor of 2 because, when considering variations, we have to include all permutations of $\mu\leftrightarrow\nu,\rho\leftrightarrow\lambda$, which results in four terms. One term is symmetric in $\mu\rho$ and can be dropped. Another term has an apparent $\varepsilon^{\mu\rho\sigma}ip_\sigma$ factor. The other two terms are not totally antisymmetric in $\mu\rho$, but after antisymmetrization they give another $\varepsilon^{\mu\rho\sigma}ip_\sigma$ factor. Therefore, we get a prefactor of $c/(48\pi)$.

Matching Eq.~\eqref{eq:Pi_ASres} with Eqs.~\eqref{eq:CSgh} and \eqref{eq:TT=ddCSg}, we see that the MCS theory has central charge $c=-\sgn(m_t)=-\sgn(k)$.

\subsection{Finite $T$: Thermal Hall Effect}
We now evaluate Eq.~\eqref{eq:Pi_AS} at nonzero $T$. Since we are interested in the thermal Hall effect, we will restrict ourselves to the energy-current sector $\nu=\lambda=0$.

Following some algebra using Mathematica, we find that the $(p+q)^2$ and $q^2$ factor in the denominator of $\Pi_{\rm AS}$ cancels out (we only write down the $(\mu0;\rho0)$ component here, but the cancellation happens for all components):
\begin{equation}\label{eq:Pi_AS=u}
    \Pi_\text{AS}^{\mu0;\rho0}(p)=\int\frac{\rd^3q}{(2\pi)^3}\frac{-m_t\,\varepsilon^{\mu\rho\sigma}u_\sigma(p,q)}{(q^2-m_t^2)((p+q)^2-m_t^2)},
\end{equation}
where $u_\sigma(p,q)$ are three polynomials in $p,q$ (superscripts denote component, not square):
\begin{widetext}
\begin{alignat}{1}
        u^0&=\frac{1}{2} \left[\left(p^1\right)^2 q^0+p^1 q^1 \left(p^0+2 q^0\right)+p^2 \left(p^2 q^0+q^2 \left(p^0+2 q^0\right)\right)\right],\\
        \nonumber
        u^1&=\frac{1}{2} \left[\left(p^0\right)^2 \left(-q^1\right)+p^0 q^0 \left(p^1-2 q^1\right)+2 \left(p^1\right)^2 q^1+p^2 q^1 \left(p^2+2 q^2\right)+p^1 \left(p^2 q^2+2 \left(q^0\right)^2+2 \left(q^1\right)^2\right)\right],\\
        \nonumber
        u^2&=\frac{1}{2} \left[\left(p^0\right)^2 \left(-q^2\right)+p^0 q^0 \left(p^2-2 q^2\right)+2 \left(p^2\right)^2 q^2+p^1 q^2 \left(p^1+2 q^1\right)+p^2 \left(p^1 q^1+2 \left(q^0\right)^2+2 \left(q^2\right)^2\right)\right].
\end{alignat}
\end{widetext}
To proceed, we note that $\Pi_\text{AS}^{\mu\nu;\rho\lambda}$ satisfies the Ward identity from both sides, so we have the ansatz
\begin{equation}
    \Pi_\text{AS}^{\mu0;\rho0}(p)=m_t\,\varepsilon^{\mu\rho\sigma}p_\sigma A(p),
\end{equation}
and
\begin{equation}
    A(p)=\frac{1}{p^2}\int\frac{\rd^3 q}{(2\pi)^3}\frac{-p\cdot u}{(q^2-m_t^2)((p+q)^2-m_t^2)}.
\end{equation}

We then evaluate the finite temperature part of $A(p)$, by replacing the frequency integral with a Matsubara summation
$$
  \int\frac{\rd q^0}{2\pi}\to iT \sum_{q^0=2\pi i Tn}.
$$
The summation can be performed by standard contour methods; the finite-temperature part is
\begin{widetext}
\begin{alignat}{1}
\nonumber
A_\beta(p)&=\frac{i}{p^2}\int\frac{\rd^2\vect{q}}{(2\pi)^2}\frac{1}{\left(E_{\vect{p}+\vect{q}}-E_{\vect q}-p^0\right) \left(E_{\vect{p}+\vect{q}}+E_{\vect q}-p^0\right) \left(E_{\vect{p}+\vect{q}}-E_{\vect q}+p^0\right) \left(E_{\vect{p}+\vect{q}}+E_{\vect q}+p^0\right)}\\
\nonumber
&\times\left\{\frac{ n_B(E_{\vect{p}+\vect{q}})}{E_{\vect{p}+\vect{q}}} \left[-\vect{p}^2E_{\vect{p}+\vect{q}}^4 +E_{\vect{p}+\vect{q}}^2 \left(E_{\vect q}^2 \vect{p}^2+\left(p^0\right)^2 (3 \vect{p}\cdot\vect{q}+2 \vect{p}^2)-\vect{p}\cdot\vect{q} (\vect{p}\cdot\vect{q}+\vect{p}^2)\right)\right.\right.\\
\nonumber
&\left.+(\vect{p}\cdot\vect{q}+\vect{p}^2) \left(E_{\vect q}^2-(p^0)^2\right)\left(\left(p^0\right)^2+\vect{p}\cdot\vect{q}\right)\right]\\
&-\frac{n_B(E_{\vect q})}{E_{\vect q}} \left[\left(\vect{p}^2 \left(E_{\vect q}^2+\vect{p}\cdot\vect{q}\right)+\left(\vect{p}\cdot\vect{q}\right)^2 \right)\left(-E_{\vect{p}+\vect{q}}^2+E_{\vect q}^2+\left(p^0\right)^2\right)\right.
+\left.\left.\vect{p}\cdot\vect{q}\left(p^0\right)^2 \left(E_{\vect{p}+\vect{q}}^2+3 E_{\vect q}^2-\left(p^0\right)^2\right)\right]\right\},
\label{eq:Abeta}
\end{alignat}
\end{widetext}
where we have dropped the zero-temperature contribution.

    To get the thermal Hall conductivity, we need to compute the Kubo conductivity $\kappa_{xy}^\text{Kubo}$ and the heat magnetization $M_Q$. As discussed earlier, because the gravitational Chern-Simons term has three derivatives, it does not contribute to $\kappa_{xy}^\text{Kubo}$ or $M_Q$, so we only need to consider the finite-temperature contributions. By inspecting \eqref{eq:Abeta}, we see that $A_\beta(p^0,\vect{p}=0)=0$, and therefore, $\kappa_{xy}^\text{Kubo}=0$.

\begin{figure}
  \centering
  \includegraphics[width=\linewidth]{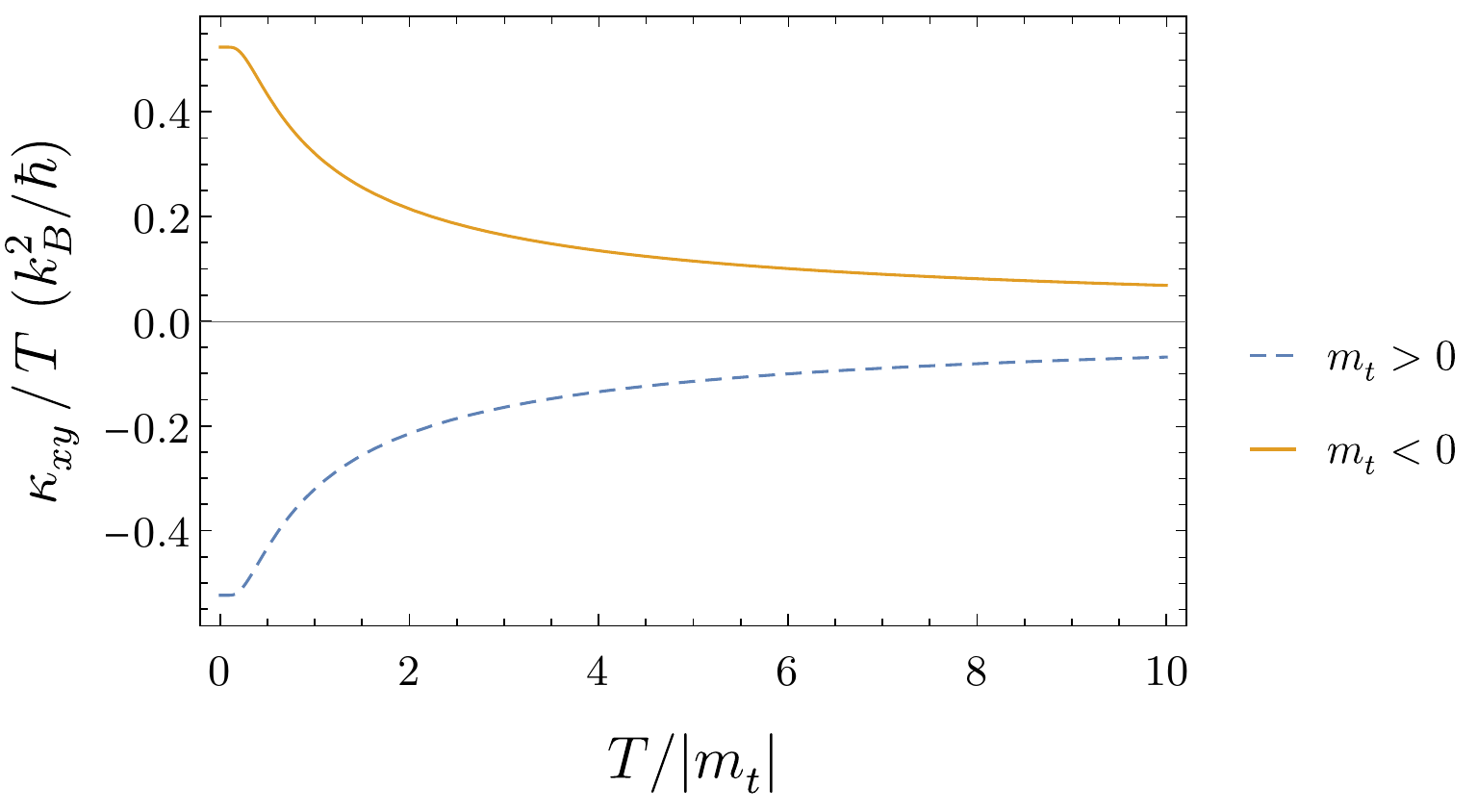}
  \caption{The thermal Hall conductivity due to the gauge fields, from Eq.~\eqref{eq:kappaxy_gauge}. $m_t$ is the topological mass. 
  }\label{fig:kappaxy2}
\end{figure}

Sending $p^0,\vect{p}\to 0$ in the static limit, we get
\begin{equation}\label{eq:Ap}
\begin{split}
  &A(p^0=0,\vect{p}\to 0)=\\
  &\frac{-i}{4\pi}\left[|m_t|n_B(|m_t|)-T\ln(1-e^{-|m_t|/T})\right].
\end{split}
\end{equation}
The heat magnetization can be obtained from the differential equation
\begin{alignat}{1}
\nonumber
    2M_Q-T\frac{\partial M_Q}{\partial T}&=-\frac{1}{2i}\nabla_{\vect p}\times \Pi_\text{AS}^{i0;00}(p^0=0,\vect{p}\to 0)\\
    &=-im_tA(p^0=0,\vect{p}\to 0),
\end{alignat} where the different prefactor compared to Eq.~\eqref{eq:rel} comes from the definition of $\Pi^{\mu\nu;\rho\lambda}$.
 Integrating the above differential equation brings us to
\begin{equation}
    \frac{M_Q}{T^2}=C(m_t)-\frac{1}{4\pi}f(m_t/T),
\end{equation}where
\begin{equation}
   f(x)=x\ln(1-e^{-|x|})-2~\sgn(x){\rm Li}_2(e^{-|x|}),
\end{equation} and the integration constant $C(m_t)$ is arbitrary function of $m_t$.
    This results in the thermal Hall conductivity
\begin{equation}\label{eq:kappaxy_gauge}
    \frac{\kappa^{}_{xy}}{T}=\frac{2M_Q}{T^2}=2C(m_t)-\frac{1}{2\pi} f(m_t/T),
\end{equation}
which is plotted in Fig.~\ref{fig:kappaxy2}.

We choose $C(m_t)=-\frac{\pi}{12}\,\sgn(m_t)$ such that $\kappa_{xy}/T$ vanishes continuously  at $m_t$\,$=$\,$k$\,$=$\,$0$. The physical motivations for this choice are the following. First, at $k=0$, we have the usual Maxwell theory, which should have no thermal Hall effect at any temperature. Secondly, in the high-temperature limit $T\gg m_t$, the system should be insensitive to the ground-state energy gap ($\sim m_t$) and the related topological distinctions, so the thermal Hall coefficient should also vanish. Therefore, $\kappa_{xy}/T$ should disappear continuously at $m_t=k=0$ and we fix $C(m_t)$ accordingly.

In the $m_t\to\infty$ limit, we obtain
\begin{equation}
    \kappa^{}_{xy}/T=-\frac{\pi}{6}\sgn(m_t)=-\frac{\pi}{6}\sgn(k),
    \label{k1}
\end{equation}
which, once again, yields central charge $c=-\sgn(m_t)=-\sgn(k)$.

\section{Dirac Chern-Simons Theory}
\label{sec:dcs}

This section turns to a discussion of the thermal Hall response of the SU(2) Dirac Chern-Simons theory in Eq.~(\ref{eq:L_CS}) in the $1/N_f$ expansion. In order to sidestep subtle issues with gravitational anomalies, we will view Eq.~(\ref{eq:L_CS}) as an effective theory, in which the Chern-Simons term is obtained by integrating out spectator heavy Dirac fermions. To obtain a SU(2) Chern-Simons term at level $k$, we need $2 |k|$ flavors of heavy Dirac fermions with mass $M$ obeying $\mbox{sgn}(M) = \mbox{sgn}(k)$. We will always assume $|M| \gg T$, while the ratio of the light Dirac fermion to temperature, $m/T$, can be arbitrary.

\begin{figure}[tb]
\includegraphics[width=\linewidth]{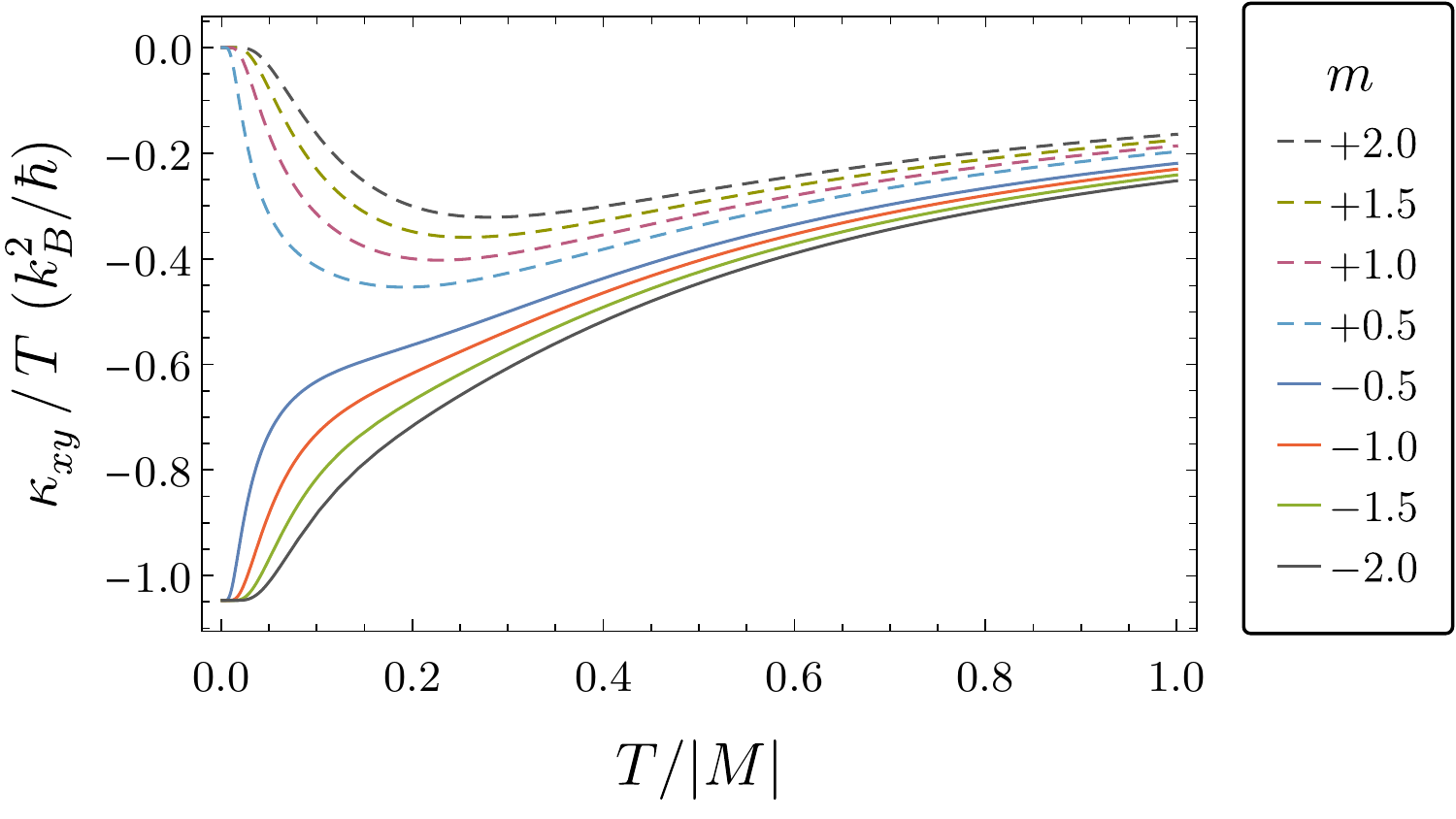}
\caption{The thermal hall conductivity computed from Eq.~\eqref{kxymMh} upon including
both light and heavy fermions of masses $m$ and $M = -10$, respectively for $N_f = 2 |k| = 1$. The quantized value in the topological phase at zero temperature is $\pi/3$, as expected from mean-field theory. The temperature
dependence of $\kappa_{xy}/T$ calculated in this continuum field theory is in excellent agreement with the results on the lattice model in Ref.~\onlinecite{THNP}}
\label{fig:Kxy2}
\end{figure}

At leading order for large $N_f$, we ignore gauge fluctuations, and simply add the contributions of the light and heavy Dirac fermions, using the results in Section~\ref{sec:dirac}.
Owing to the SU(2) gauge index carried by the $\Psi$ fermions in Eq.~(\ref{eq:L_CS}), we need to multiply the contribution by an additional factor of 2 for each flavor. In this manner, we obtain the thermal Hall conductivity
\beq
\kappa^{}_{xy} = 2 N^{}_f \, \kappa^{}_{xy,D} (m) + 4 |k| \, \kappa^{}_{xy,D} (M)
    \label{kxymMh}
\eeq
where $\kappa_{xy,D} (m)$ is the Dirac fermion contribution in Eq.~(\ref{eq:kxy_gt}).
We show a plot of Eq.~(\ref{kxymMh}) in Fig.~\ref{fig:Kxy2} for $N_f$\,$=$\,$2\lvert k \rvert$\,$=$\,1. In the limit $|M|/T \rightarrow \infty$ and $|m|/T \rightarrow \infty$, Eq.~(\ref{kxymMh}) yields the first term in the square brackets in Eq.~(\ref{kappa_exact}).

Upon examining the effect of gauge fluctuations in the $1/N_f$ expansion, we find that there are Feynman graphs which potentially contribute to the thermal Hall conductivity even at $N_f = \infty$. However, evaluation of these graphs shows that they vanish, as we will illustrate in Section~\ref{sec:Ninfinity}, so no corrections are needed to Eq.~(\ref{kxymMh}) at this order. We then discuss the leading $1/N_f$ corrections in Section~\ref{sec:1N}.

\subsection{Gauge fluctuations in the $N_f\rightarrow \infty$ limit}
\label{sec:Ninfinity}

Explicitly expanding out the non-Abelian gauge field, the appropriate modification of the Lagrangian~\eqref{eq:L_CS}, for a particular fermion species, reads as
\begin{alignat}{1}
\nonumber
\mathcal{L}_{N_f, SU(2)}^\pdagger &= \sum_{v = 1}^{N_f}  \sum_{b,s,t} i \bar{\Psi}^\pdagger_{v s} \gamma^\mu \left(\partial_\mu - \frac{i}{\sqrt{N_f}} a_\mu^b \tau^b_{s t}  \right) \Psi_{v t}^\pdagger \\&+ m \bar{\Psi}^\pdagger_{v s} \Psi_{v s}^\pdagger,
\label{eq:LSU(2)}
\end{alignat}
where $\tau^a$ are the generators of SU(2), $A_\mu = a_\mu^b \tau^b$, and the coupling constant has been scaled by $1/\sqrt{N_f}$ for normalization. The fermionic field is labeled simultaneously by the flavor index $v = 1,\ldots,N_f$ as well as the color index $i = 1, 2, 3$; the $\gamma$ (or $\sigma$) and $\tau$ Pauli matrices operate in Dirac and color spaces, respectively. The fermion-gluon vertex corresponds to
\begin{equation}
\includegraphics[width=0.7\linewidth]{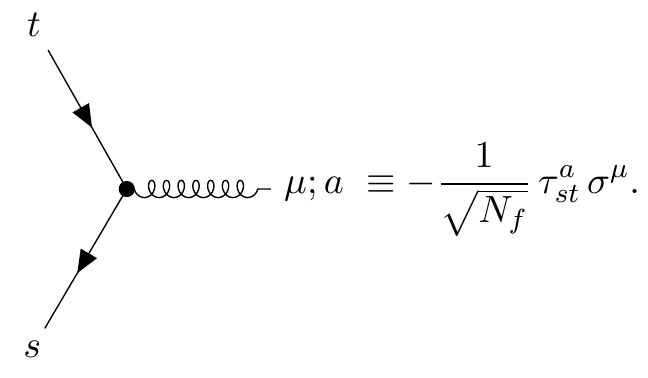}
\end{equation}

\begin{figure}[tb]
\includegraphics[width=\linewidth,trim={0.25cm 0 0.25cm 0},clip]{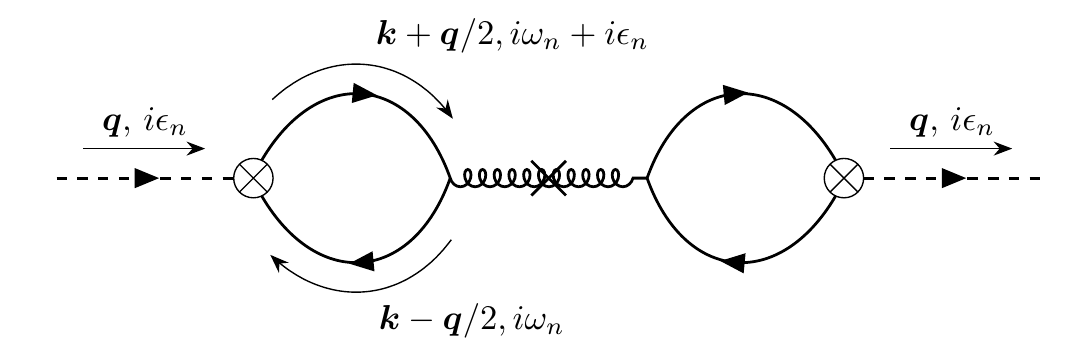}
\caption{\label{fig:gauge} The sole Feynman diagram contributing at leading order to gauge-field corrections in the limit $N_f \rightarrow \infty$. The dressed gluon propagator, indicated by a cross, is obtained by perturbatively integrating out the fermions from the action $\mathcal{S}$.}
\end{figure}

Every fermion loop now bears an extra factor of $N_f$ owing to the summation over flavors, while each interaction vertex carries a factor of $1/\sqrt{N_f}$. The diagram in Fig.~\ref{fig:fbubble} is, therefore, of $\mathcal{O}(N_f)$. In the limit of large $N_f \rightarrow \infty$, the only diagram that contributes at the same order is shown in Fig.~\ref{fig:gauge}.

\noindent One might naively think that there are additional diagrams beyond Fig.~\ref{fig:gauge} because each new fermion bubble \textit{within} the gluon propagator is of $\mathcal{O}(1)$ in this expansion. Typically, these contributions can be subsumed in a renormalized propagator, denoted by a cross, by summing up the chain of bubble diagrams in a geometric series as
\begin{equation}
\includegraphics[width=\linewidth,trim={0.2cm 0 0.1cm 0},clip]{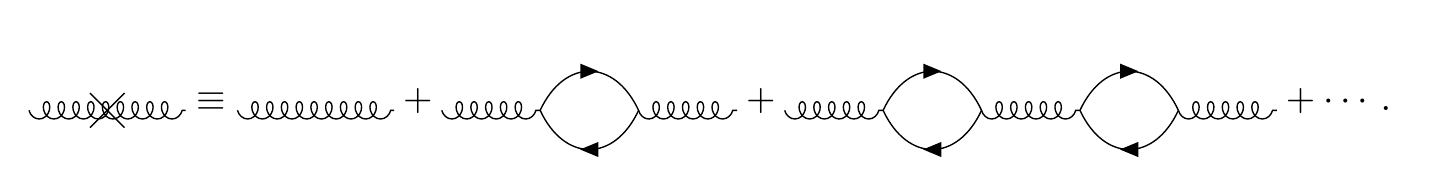}
\end{equation}
However, in 2+1D the Maxwell kinetic term is irrelevant and the bare $(F_{\mu \nu}^a)^2$ is thus suppressed by higher energy scales. Indeed, in our formulation, the bare kinetic term comes from integrating out the heavy fermions, and is proportional to $1/M$. The renormalized gluon propagator should also include the contribution from integrating out light fermions, and in fact, it is dominated by the light fermion bubble.  To derive the renormalized gluon propagator, we build upon the results of Ref.~\onlinecite{kaul2008quantum} for the photon propagator at nonzero temperatures in 2+1D U(1) gauge theories with fermionic and bosonic matter. The full expression for the gluon propagator is detailed in Appendix~\ref{app:gluon}.

\begin{figure}[t]
\includegraphics[width=\linewidth,trim={0.25cm 0 0.25cm 0},clip]{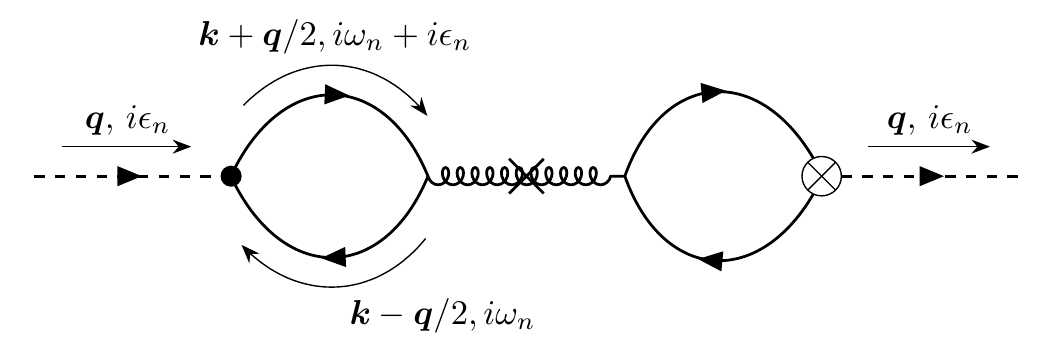}
\caption{\label{fig:gauge2} The magnetization diagram at the same order in $1/N_f$ as Fig.~\ref{fig:gauge}. One of the two thermal vertices in Fig.~\ref{fig:gauge} is replaced by an energy-density vertex here.}
\end{figure}

It is now easy to observe the absence of gauge-field corrections at leading order. Figure~\ref{fig:gauge} is composed of two fermion bubbles, each of which, following the Feynman rules listed earlier, translate to
\begin{equation}
\label{eq:vanish1}
\begin{split}
&\frac{T}{V} \sum_{\vect{k}, i\omega_n, s\, t} \left(\frac{- \tau^{a}_{st}}{\sqrt{N_f}}\right)\delta^{st}\, (i \omega_n + i\epsilon_n/2)\\
&\mathrm{Tr} \left[\mathcal{G} (\vect{k}+\vect{q}/2, i \omega_n + i \epsilon_n)\, \sigma^\nu \,\mathcal{G} (\vect{k}-\vect{q}/2, i \omega_n)\, \sigma^\mu \right],
\end{split}
\end{equation}
where the factor of $\delta^{st}$ comes from the fact that the thermal vertex conserves color. Resultantly, Eq.~\eqref{eq:vanish1} is just proportional to $\mbox{Tr}\,(\tau^a)$ and hence, is identically zero. By the same reasoning, the diagram for the magnetization contribution in Fig.~\ref{fig:gauge2} also vanishes. Therefore, we conclude that upon taking $N_f \rightarrow \infty$, there are no corrections to the thermal Hall conductivity due to SU(2) gauge-field fluctuations.

\subsection{Gauge fluctuations at next-to-leading order}
\label{sec:1N}

We are now positioned to consider the contributions to $\kappa_{xy}$ of the theory (\ref{eq:L_CS}) at order $N_f^0$. The Feynman diagrams which contribute to this order are shown in Figs.~\ref{fig:all_diagrams} and \ref{fig:Aslamazov_Larkin}. Given the complexity of the gluon propagator in Appendix~\ref{app:gluon},
and the differential equation that has to be solved for the magnetization subtraction, we do not attempt a full numerical evaluation of these graphs for general $m/T$. Instead, we will be satisfied by examining them in the limit $|m|/T \rightarrow \infty$ (recall that we always take the limit $|M|/T \rightarrow \infty$). In this limit, we expect that a description in terms of the effective Maxwell-Chern-Simons theory in Section~\ref{sec:mcs} applies, and we can therefore deduce the contribution to $\kappa_{xy}$ from results therein. The MCS theory gives the second term in the bracket of Eq.~\eqref{kappa_exact}.

\begin{figure}[htb]
    \centering
    \includegraphics[width=\linewidth]{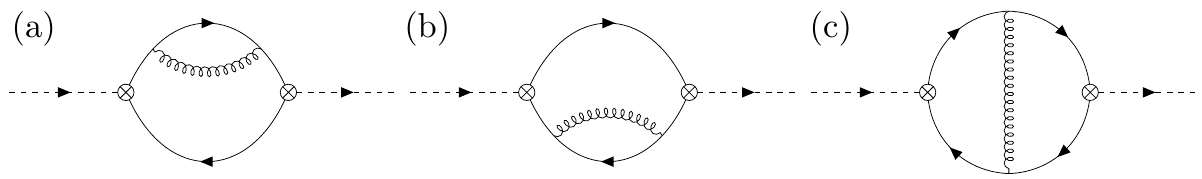}
    \caption{The (a--b) density of states (DOS) and (c) Maki-Thompson \cite{maki1968critical, thompson1970microwave} diagrams, which contribute to $\kappa_{xy}$ for the theory (\ref{eq:L_CS}) at $\mathcal{O}(N_f^0)$. Additionally, the magnetization subtraction requires evaluation of the analogous graphs given by replacing a thermal vertex with an energy-density vertex, like in Fig.~\ref{fig:gauge2}.}
    \label{fig:all_diagrams}
\end{figure}


 We now argue that the DOS and MT diagrams listed in Fig.~\ref{fig:all_diagrams} are not important in the $|m|/T\to\infty$ limit. The DOS diagram is simply adding self-energy into the fermion propagator. A standard computation yields a fermion mass correction $\delta m\propto e^2/N_f$ in the zero-momentum limit, where $e^2$ is the coupling constant of the gauge field. The DOS diagram can also generate fermion anomalous dimension at higher momentum. The MT diagram is a vertex correction to the stress tensor. Using a gravitational Ward identity in \cite{GravitationalWardIdentity}, it can be shown that the anomalous dimension from the vertex correction cancels that from the self energy, and the net effect is a fermion mass renormalization $\delta m$ consistent with the self-energy calculation. We note in passing that similar behaviors have been observed in nonrelativistic calculations \cite{FluctuatingSuperconductivity}. As a result, a finite renormalization $\delta m$ (also subleading in large $N_f$) can be ignored in the $|m|/T\to\infty$ limit.

 Therefore, the important Feynman diagram in this limit is the `Aslamazov-Larkin' diagram \cite{aslamazov1968effect} drawn in Fig.~\ref{fig:Aslamazov_Larkin}. The triangular vertices in Fig.~\ref{fig:Aslamazov_Larkin} each reduce to the stress-energy vertex used in Section~\ref{sec:mcs}, as we now show.

\begin{figure}[htb!]
\centering
\includegraphics[width=0.5\linewidth]{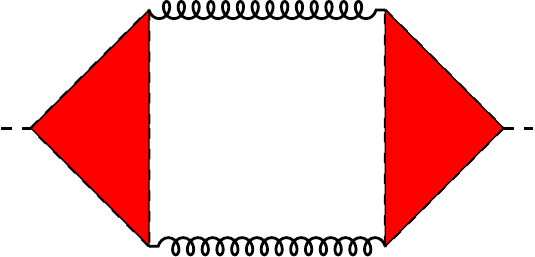}

\caption{\label{fig:Aslamazov_Larkin} The `Aslamazov-Larkin' diagram responsible for the thermal Hall response at order $N_f^0$. The red triangles denote the effective stress tensor-gauge field-gauge field vertex obtained from integrating out fermionic loops in Fig.~\ref{fig:triangular_vertex}.}
\end{figure}

Following the discussions of Appendix~\ref{app:feynmanrules}, the gauge-invariant stress tensor of the theory \eqref{eq:LSU(2)} is
\begin{equation}\label{}
\begin{split}
  T^{\mu\nu}&= \frac{i}{2}\bar{\Psi}\gamma^\mu(\overrightarrow{\partial}^\nu-\overleftarrow{\partial}^\nu)\Psi+\frac{a^{\nu b}}{\sqrt{N_f}}\bar{\Psi} \gamma^\mu \tau^b \Psi\\&-\eta^{\mu\nu}\mathcal{L}_{N_f,SU(2)}.
\end{split}
\end{equation} For shorter notation, we have suppressed flavor and color indices on the fermions.

\begin{figure}[htb]
\centering
\includegraphics[width=0.8\linewidth]{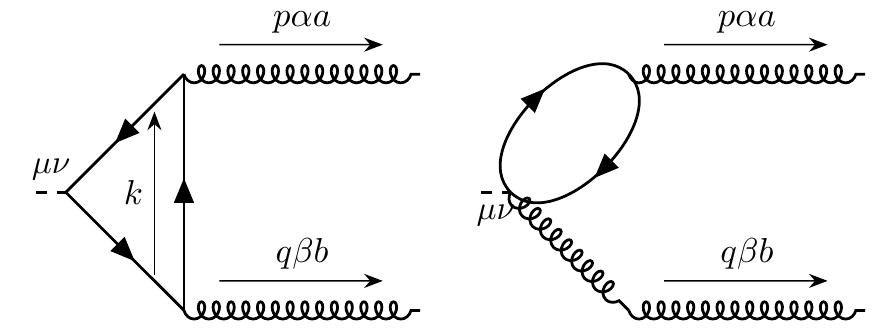}
\caption{\label{fig:triangular_vertex} The two types of diagrams for the triangular vertex.}
\end{figure}

Based on the above stress tensor, there are two types of vertices contributing to the triangular vertex, as shown in Fig.~\ref{fig:triangular_vertex}.
\begin{widetext}

The first type is a fermion triangle; the corresponding effective vertex function is
\begin{equation}\label{eq:gamma1}
\begin{split}
  \Gamma_1^{\mu\nu\alpha\beta}(p,q)\frac{\delta^{ab}}{2}&=\int\frac{\rd^3 k}{(2\pi)^3}(-1){\rm Tr}\left\{\left[\gamma^\mu\frac{(2k+q-p)^\nu}{2}-\eta^{\mu\nu}(\frac{2\slashed{k}+\slashed{q}-\slashed{p}}{2}+m)\right]\right. \\
  &\times\left.\frac{i}{\slashed{k}-\slashed{p}+m}i\gamma^\alpha\tau^a\frac{i}{\slashed{k}+m}
  i\gamma^\beta\tau^b\frac{i}{\slashed{k}+\slashed{q}+m}
  \right\}+\left(p\alpha a\leftrightarrow q\beta b\right).
\end{split}
\end{equation}
The second type is a fermion bubble, and the associated effective vertex is
\begin{equation}\label{eq:gamma2}
  \begin{split}
     \Gamma_2^{\mu\nu\alpha\beta}(p,q)\frac{\delta^{ab}}{2} & =\int\frac{\rd^3 k}{(2\pi)^3}(-1){\rm Tr}\left\{\tau^b(\gamma^\mu\eta^{\nu\beta}-\gamma^\beta\eta^{\mu\nu})\frac{i}{\slashed{k}+m}i\gamma^\alpha\tau^a \frac{i}{\slashed{k}+\slashed{p}+m}\right\}\\
     &+\left(p\alpha a\leftrightarrow q\beta b\right).
  \end{split}
\end{equation}
\end{widetext}
In the equations above, we have factored out the color indices on the LHS. Since we are looking at the $|m|/T\to\infty$ limit, we will only evaluate the above integrals at zero temperature. The integrals can be performed with the standard Feynman parameter tricks. While it is possible to obtain closed-form results for arbitrary momenta and mass, the resultant expressions are too long and not very enlightening. We expand the result to second order in momenta, and obtain
\begin{equation}\label{}
  \Gamma_1^{\mu\nu\alpha\beta}(p,q)+\Gamma_2^{\mu\nu\alpha\beta}(p,q)=\frac{1}{12\pi|m|}\Gamma^{\mu\nu\alpha\beta}(p,q),
\end{equation} where $\Gamma^{\mu\nu\alpha\beta}(p,q)$ is the stress-tensor vertex function defined in \eqref{eq:Gamma}.

Given the identity of the stress tensor above, we can now use the results of Section~\ref{sec:mcs}
on the Maxwell-Chern-Simons theory to deduce the $1/N_f$ correction to the Dirac Chern-Simons theory in the limit $|m|/T \rightarrow \infty$. Each U(1) gauge field yields the contribution in (\ref{k1}); for a SU(2) gauge field, we have 3 U(1) gauge fields (which can be treated as independent at this order in $1/N_f$), so we obtain the second term in (\ref{kappa_exact}) in the limit of large $|\hat{k}|$.

\section{Conclusion}

We have examined the thermal Hall conductivity in square-lattice insulators
near the quantum phase transition
between the N\'eel state and a state with coexisting N\'eel and semion topological order. This transition is described by the Dirac Chern-Simons field theory in (\ref{eq:L_CS}) for $N_f=1$ and $k=-1/2$. The thermal Hall conductivity is expected to obey the universal scaling form in (\ref{kappauniv}). In the limit of low $T$ away from the critical point, $|m|/T \rightarrow \infty$, we have the exact result in (\ref{kappa_exact}) obtained via a sophisticated mapping to conformal field theories on the boundary of the sample. We obtained the leading and next-to-leading order results of (\ref{kappa_exact}) in a direct $1/N_f$ expansion (with $k$ taken of order $N_f$). These computations can also be applied to other values of $m/T$, and results to leading order are in (\ref{eq:kxy_gt}) and Fig.~\ref{fig:Kxy2}; however, the next-to-leading order computations are numerically demanding.

One of the lessons of this computation is that the leading contribution can be viewed as that of fermionic matter, while the next-to-leading order terms arises from the quantum fluctuations of the gauge fields (here we are viewing the Chern-Simons term in the field theory as arising from integrating out a massive fermionic matter field).

We applied this lesson to a model of the doped antiferromagnet described by (\ref{eq:Lf}). This theory contains fermionic matter forming pocket Fermi surfaces: the thermal Hall contribution of these pockets is assumed to obey the Wiedemann-Franz law. The contribution of the gauge field was deduced from the Maxwell-Chern-Simons effective action in (\ref{MCS}), which has the thermal Hall contribution specified by
(\ref{eq:kappaxy_gauge}). Importantly, this contribution has the opposite sign from the Wiedemann-Franz contribution, consistent with the experimental trends \cite{Taillefer19}.
We note that the Dirac fermion thermal Hall conductivity in (\ref{eq:kxy_gt}) and the Maxwell-Chern-Simons thermal Hall conductivity in (\ref{eq:kappaxy_gauge}) correspond to distinct universal scaling functions, a consequence of the nontopological nature of the thermal Hall effect in the quantum-critical crossover regime.

The gauge-field contribution has the correct sign to account for the additional negative contribution to $\kappa_{xy}/T$ in the pseudogap regime, as observed in Ref.~\onlinecite{Taillefer19}. However, its magnitude is bounded by $\pi/6$ [see (\ref{eq:kappaxy_gauge})] for the case a single U(1) gauge field. The observed magnitude is larger by, at least, a factor of 2; the coupling constants in (\ref{MCS}) only appear in the crossover energy scale $m_t = \sigma_{xy}/(e^2 K_2)$, and not in the overall magnitude of $\kappa_{xy}$.
It is possible that other models of the pseudogap with additional gauge fields could account for the discrepancy. Alternatively,
the phonon contribution  \cite{Kivelson19} needs to be combined with the emergent gauge field to understand the observations, and a phonon-emergent photon coupling
could provide the needed chirality in the phonon transport.

\begin{acknowledgments}
This research was supported by the National Science Foundation under Grant No. DMR-1664842.
We thank G.~Grissonnanche, A. Kapustin, and L.~Taillefer for helpful discussions.
RS acknowledges useful discussions with Holmfridur Hannesdottir and Harley Scammell.
\end{acknowledgments}

\appendix

\section{Low-energy field theory}
\label{LowEnergyFieldTheory}

The form of the continuum field theory is clear: upon inspection of the spectrum of the lattice model in \refcite{THNP} close to the phase transition, we observe two Dirac cones (of complex fermions) that result from the two sublattices of our ansatz. Therefore, the ($2\times 2$) Dirac matrices, $\gamma_\mu$, will act in sublattice space. From the lattice theory, with spinon operators $f_{i\sigma}$, $\sigma =\uparrow, \downarrow$, we know that there will be an SU(2) gauge field. The lattice gauge transformations act locally in the lattice model,
\begin{equation}
\text{SU}(2)_g: ~ \begin{pmatrix} f^\dagger_{i\downarrow} \\ f_{i\uparrow} \end{pmatrix} \, \longrightarrow \, U_g(i)\begin{pmatrix} f^\dagger_{i\downarrow} \\ f_{i\uparrow} \end{pmatrix} , \quad U_g(i)\in \text{SU}(2); \label{LatticeGaugeTrafo}
\end{equation}
as such, they cannot mix different sublattices in the continuum model. It will therefore have the Lagrangian
\begin{equation}
\mathcal{L}_{D} = i \bar{\Psi} \gamma^\mu \left(\partial_\mu - i A_\mu \right)\Psi + m \bar{\Psi}\Psi -\frac{1}{2}\mbox{\textit{CS}}[A_\mu],   \label{CompactDiracHam}
\end{equation}
with $\bar{\Psi} = \Psi^\dagger \gamma^0$. As usual, the Dirac matrices satisfy the Clifford algebra, $\{\gamma^\mu,\gamma^\nu\} = 2\eta^{\mu\nu}$ with $\eta = \text{diag}(+,-,-)$. The Chern-Simons term is an obvious consequence of the additional massive fermions and will be omitted in the following. To set up the notation, let us write \equref{CompactDiracHam} more explicitly,
\begin{align}\begin{split}
\mathcal{L}_{D} &= i \Psi^\dagger_{s\alpha} (\gamma^0\gamma^\mu)_{ss'} \left[\partial_\mu - i (A_\mu)_{\alpha\alpha'} \right]\Psi_{s'\alpha'} \\ &\qquad + m \Psi^\dagger_{s\alpha} (\gamma_0)_{ss'}\Psi_{s'\alpha},   \label{ExplicitDiracHam}
\end{split}\end{align}
with sublattice and gauge index $s$ and $\alpha$, respectively. Here, the gauge transformations act as
\begin{align}
\begin{split}
\mathrm{SU}(2)_g: \quad \Psi_{s\alpha} \, &\rightarrow \, (U_g)_{\alpha\alpha'}\Psi_{s\alpha'} ;\\
                  \quad  A_\mu \,&\rightarrow \, U^\pdagger_g A_\mu U_g^\dagger - i(\partial_\mu U^\pdagger_g) U_g^\dagger. \label{GaugeTrafoOnPsi}
\end{split}
\end{align}

In the remainder of this section, we will derive \equref{ExplicitDiracHam} from the lattice model and, thereby, relate $\Psi$ explicitly to the lattice fermions. We note that the lattice model also contains monopole operators, but we assume that these are irrelevant at the critical point \cite{Song:2018ial}.

For simplicity, let us focus on the case without a Zeeman field, $\vect{B}^{}_Z = 0$, and use the same gauge as in \refcite{THNP}. We define the Fourier transform as
\begin{equation}
\begin{split}
&f_{i\sigma} = \frac{1}{\sqrt{N}} \sum_{\vect{k}} e^{i\vect{k}\vect{x}_i} f_{\vect{k}\sigma s(i)}, \quad \vect{x}_i = (i_x,i_y), \\
  & s(i) = \begin{cases} A, \quad i_x + i_y \text{ even}, \\ B,  \quad i_x + i_y \text{ odd}.\end{cases} \label{DefinitionOfFourierTransform}
\end{split}
\end{equation}
The spectrum of the lattice model has minima at $\vect{Q}$ and $-\vect{Q}$, where $\vect{Q}=(\pi/2,0)^T$, with spin polarization $\uparrow$ and $\downarrow$, respectively. Let us expand around these minima by defining new ``slow'', low-energy fields
\begin{equation}
c_{\vect{q},s,v=+} := f_{\vect{Q}+\vect{q},\uparrow,s} ,~ c_{\vect{q},s,v=-} := f_{-\vect{Q}+\vect{q},\downarrow,s},~ |\vect{q}| \ll 1/a,
\end{equation}
for each of the two ``valleys'' $v=\pm$ and sublattices $s=A,B$. Equivalently, this corresponds to
\begin{equation}
f_{i\uparrow} \sim e^{i\vect{Q}\vect{x_i}}c_{s(i)+}(\vect{x}_i) ,~ f_{i\downarrow} \sim e^{-i\vect{Q}\vect{x_i}}c_{s(i)-}(\vect{x}_i)  \label{RealSpaceDefinitionOfc}
\end{equation}
in real space, i.e., $c_{s,v}(\vect{r}) := N^{-1/2} \sum_{\vect{q}}^\Lambda e^{i\vect{q}\vect{r}}c_{\vect{q},s,v}$ with some cutoff $\Lambda \ll 1/a$. With these definitions, the mean-field Hamiltonian can be written as
\begin{equation}
\begin{split}
H_{\text{MF}} \sim &- i v_F \int\diff \vect{r} \, c^\dagger_{s,v} \left[ v (\tau_x)_{ss'}\partial_x + (\tau_y)_{ss'} \partial_y \right] c^\pdagger_{s',v} \\
&+  m \int\diff \vect{r} \, c^\dagger_{s,v} v(\tau_z)_{ss'} c^\pdagger_{s',v} \label{MFHamiltonian}
\end{split}
\end{equation}
at low-energies. Here $v_F = 2t_1$ (which we will set to $1$ in the following) and $m=-(4t_2+N_z/2)$, where, as in \refcite{THNP}, $t_1$, $t_2$, and $N_z$ are the nearest-, next-nearest-neighbor hopping, and the N\'eel order parameter, respectively; furthermore, $\tau_j$ denote Pauli matrices in sublattice space.

As follows from comparison of \equsref{LatticeGaugeTrafo}{RealSpaceDefinitionOfc}, gauge transformation act as
\begin{equation}
\mathrm{SU(2)}_g: ~ C_s(\vect{r})\,\rightarrow \, U^{(s)}_g(\vect{r}) C_s(\vect{r}), ~ C_s(\vect{r}):=\begin{pmatrix} c_{s,-}^\dagger(\vect{r}) \\ c_{s,+}(\vect{r}) \end{pmatrix}
 \end{equation}
in the low-energy theory. Naively, one might think that the gauge transformations in the two sublattices are independent as they were in the lattice model. However, this is would enhance the gauge symmetry to SU(2) $\times$ SU(2)  in the continuum model which is not the case; the reason is that not all gauge transformations allowed on the lattice act entirely in the low-energy field theory. In fact, we will see that $U^{A}_g(\vect{r})$ and $U^{B}_g(\vect{r})$ are related by a similarity transformation within the continuum field theory,
\begin{equation}
U^{A}_g(\vect{r}) = V^\dagger U^{B}_g(\vect{r}) V, \qquad V \in \text{SU}(2). \label{UnitaryTrafo}
\end{equation}
This means that there is a gauge, reached by performing a gauge transformation with $U_g(i) = V^\dagger$ for $i_x+i_y$ even and $U_g(i) = \mathbbm{1}$ for $i_x+i_y$ odd, where the gauge transformation is independent of $s$ in the continuum model. To see this, just note that the new field after the gauge transformation,
\begin{equation}
	\widetilde{C}_s = V_s C_s, \qquad V^A = V^\dagger, \quad V^B = \mathbbm{1}, \label{CTilde}
\end{equation}
transforms as $\widetilde{C}_s(\vect{r}) \rightarrow V_s^\dagger U^{(s)}_g(\vect{r}) V_s \widetilde{C}_s(\vect{r})$; with the above choice for $V_s$, it holds that $V_s^\dagger U^{(s)}_g(\vect{r}) V_s = U^{B}_g(\vect{r})$, independent of $s$.

To add gauge-field fluctuations to the mean-field Hamiltonian (\ref{MFHamiltonian}), we rewrite the latter in terms of the Nambu field $C_s$. Denoting Pauli matrices in Nambu/valley space by $\eta_j$, we get
\begin{equation}
\begin{split}
 H_{\text{MF}} &\sim  i  \int\diff \vect{r} \, C^\dagger_{s} \left[(\tau_x)_{ss'}\partial_x + (\tau_y)_{ss'} \partial_y \right]\eta_z C^\pdagger_{s'} \\
 &+  m \int\diff \vect{r} \, C^\dagger_{s} (\tau_z)_{ss'}\eta_0 C^\pdagger_{s'}.
 \end{split}
 \end{equation}
To bring the theory to the form of \equref{ExplicitDiracHam}, we have to transform $\eta_z$ into $\eta_0$ in the first term. This can be done by introduction of a new field $\widetilde{C}_s$ as defined in \equref{CTilde} with $V=-i\eta_3$; we get
\begin{equation}
\begin{split}
  H_{\text{MF}} &\sim  i \int\diff \vect{r} \, \widetilde{C}^\dagger_{s} \left[(\tau_y)_{ss'}\partial_x - (\tau_x)_{ss'} \partial_y \right]\eta_0 \widetilde{C}^\pdagger_{s'} \\&+  m \int\diff \vect{r} \, \widetilde{C}^\dagger_{s} (\tau_z)_{ss'}\eta_0 \widetilde{C}^\pdagger_{s'}.
\end{split}
 \end{equation}
In this form, it becomes apparent that only gauge transformations, $\widetilde{C}_s(\vect{r}) \rightarrow \widetilde{U}^{(s)}_g(\vect{r})  \widetilde{C}_s(\vect{r})$, that are independent of $s$, $\widetilde{U}^{(s)}_g(\vect{r}) = \widetilde{U}_g(\vect{r})$, can appear in the low-energy theory. Based on our discussion above, we see that this indeed corresponds to \equref{UnitaryTrafo} with $V=-i\eta_3$.

To match the common conventions for the Dirac matrices,
\begin{equation}
(\gamma^0,\gamma^x,\gamma^y) = (\tau_y,i\tau_z,i\tau_x),
\end{equation}
we perform yet another unitary transformation in sublattice space only (which, thus, does not affect the gauge transformation properties),
\begin{equation}
\Psi_s(\vect{r}) = \left(e^{i\frac{\pi}{4}\tau_x}e^{i\frac{\pi}{4}\tau_z}\right)_{ss'} \widetilde{C}_{s'}(\vect{r}), \label{PsiDef}
\end{equation}
leading to
\begin{equation}
\begin{split}
  H_{\text{MF}} &\sim  i \int\diff \vect{r} \, \Psi^\dagger_{s} \left[(\tau_x)_{ss'}\partial_x - (\tau_z)_{ss'} \partial_y \right]\Psi^\pdagger_{s'} \\&+  m \int\diff \vect{r} \, \Psi^\dagger_{s} (\tau_y)_{ss'} \Psi^\pdagger_{s'}.\label{FinalMFHam}
\end{split}
\end{equation}
As $\Psi_{s\alpha}(\vect{r}) \rightarrow (U_g(\vect{r}))_{\alpha\alpha'}\Psi_{s\alpha'}(\vect{r})$ under gauge transformations, adding gauge fluctuations to \equref{FinalMFHam} in the action formalism leads precisely to \equref{ExplicitDiracHam}. We now know the relation between the field $\Psi$ and the lattice degrees of freedom.

\section{Derivation of heat current and the associated Feynman rules}
\label{app:feynmanrules}

Since our eventual interest lies in the (thermal) current-current correlation functions, our first step is to compute the associated heat current vertex. There are two approaches as described below.

\subsection{Equation-of-motion approach}
To develop the formalism, we begin with a generalized Hubbard-like model on a lattice described by the Hamiltonian
\begin{equation}
\begin{split}
    \mathcal{H} &= \sum_{i j \mu \nu} t_{ij}^{\mu \nu}\, \psi^\dagger _{i \mu} \psi^{\pdagger}_{j \nu}\equiv \sum_i h^{\pdagger}_i; \\ h^{\pdagger}_i &= \frac{1}{2} \sum_{j \mu \nu} \left(t^{\mu \nu}_{ij}\, \psi^\dagger _{i \mu} \psi^{\pdagger}_{j \nu} +t^{\nu \mu }_{ji} \, \psi^{\dagger}_{j \nu} \psi^{\pdagger}_{i \mu} \right),
\end{split}
\end{equation}
for which we shall evaluate the thermal current using the equation-of-motion technique \cite{mahan2000many}. We emphasize that $\mathcal{H}$ and the $\psi$ fermions need not be the same as those of the lattice model in \refcite{THNP}; in practice, we extract $\mathcal{H}$ solely from the effective theories such as  Eq.~\eqref{eq:action} below. In the equation above, $h_i$ stands for the local energy density; $i, j$ are the lattice sites whereas $\mu, \nu$ connote any other degrees of freedom. It readily follows \cite{paul2003thermal} that
\begin{equation}
\begin{split}
&\dot{h}^{\pdagger}_i = \frac{1}{2} \sum_{j \mu \nu} \\
&\left[t^{\mu \nu}_{ij} \left(\psi^\dagger _{i \mu} \dot{\psi}^{\pdagger}_{j \nu}-\dot{\psi}^\dagger _{i \mu} \psi^{\pdagger}_{j \nu} \right)+t^{\nu \mu }_{ij}  \left( \dot{\psi}^{\dagger}_{j \nu} \psi^{\pdagger}_{i \mu} - \psi^{\dagger}_{j \nu} \dot{\psi}^{\pdagger}_{i \mu}  \right)\right],
\end{split}
\end{equation}
where $\dot{\mathcal{O}} = i [\mathcal{H}, \mathcal{O}]$. From the continuity equation for the energy current \cite{cooper1997thermoelectric, michaeli2009quantum,Kapustin19}, $\dot{h}_i +\nabla \cdot \vect{J}^\textsc{e}_i = 0$, we find
\begin{alignat}{2}
\nonumber
\vect{J}_{\vect{q}}^\textsc{e} &= \frac{1}{\sqrt{V}}\sum_i e^{-i \vect{q}\cdot\vect{r}_i}  \vect{J}^\textsc{e}_i && \\
\nonumber
&= \frac{i}{2 \sqrt{V}} \sum_{\vect{k} \mu \nu}\,\, \frac{\partial h^{\mu \nu}_{\vect{k}}}{\partial \vect{k}} &&\bigg(\psi^\dagger_{\vect{k} - \vect{q}/2, \mu} \, \dot{\psi}^{\pdagger}_{\vect{k} + \vect{q}/2, \nu} \\
&  &&- \dot{\psi}^\dagger_{\vect{k} - \vect{q}/2, \mu} \, \psi^{\pdagger}_{\vect{k} + \vect{q}/2, \nu}  \bigg),
\label{eq:cont}
\end{alignat}
where $h^{\mu \nu}_{\vect{k}}$ is the second-quantized Hamiltonian and we have used the approximation $h_{\vect{k} + \vect{q}/2} - h_{\vect{k} - \vect{q}/2} \approx  (\partial h_{\vect{k}} /\partial \vect{k})\cdot \vect{q}$, concentrating on the small $\vect{q}$ limit.
Using the Heisenberg equation of motion, this simplifies to
\begin{equation}
\begin{split}
\vect{J}_{\vect{q}}^\textsc{e} &= -\frac{1}{2 \sqrt{V}} \sum_{\vect{k}, \mu \nu \rho} \left(\frac{\partial h^{\mu \rho}_{\vect{k}}}{\partial \vect{k}}  \,h^{\rho \nu }_{\vect{k}+\vect{q}/2} +  h^{\mu \rho}_{\vect{k}-\vect{q}/2}\, \frac{\partial h^{\rho \nu }_{\vect{k}}}{\partial \vect{k}}  \right)\\
&\times{\psi}^\dagger_{\vect{k} - \vect{q}/2, \mu}\,  \psi^{\pdagger}_{\vect{k} + \vect{q}/2, \nu}.
\end{split}
\end{equation}
In this notation, it is clear that the indices $\mu, \nu$ keep track of the component of the Dirac fermion under consideration. The heat current ($\vect{J}^\textsc{q}$) is related to the energy current by $ \vect{J}^\textsc{q} =  \vect{J}^\textsc{e}  - \mu \vect{J} $. Switching to frequency domain from Eq.~\eqref{eq:cont}, at $\mu = 0$, we obtain,
\begin{equation}
\begin{split}
\vect{J}^\textsc{q} (\vect{q}, i \epsilon_n) &= \frac{1}{\beta \sqrt{V}} \sum_{\vect{k}, i \omega_n} \left (\partial^\pdagger_{\vect{k}} h_{\vect{k}}^{\mu \nu} \right) \left(i \omega_n + i \epsilon_n/2\right) \\
&\times\psi^\dagger_{\vect{k}-\vect{q}/2, \mu} (i \omega_n)  \,\psi^\pdagger_{\vect{k}+\vect{q}/2, \nu} (i \omega_n+ i \epsilon_n),
\end{split}
\end{equation}
where $\omega_n$ and $\epsilon_n$ are fermionic and bosonic Matsubara frequencies, respectively. This defines the heat/energy-current vertex
\begin{equation}
\includegraphics[width=\linewidth]{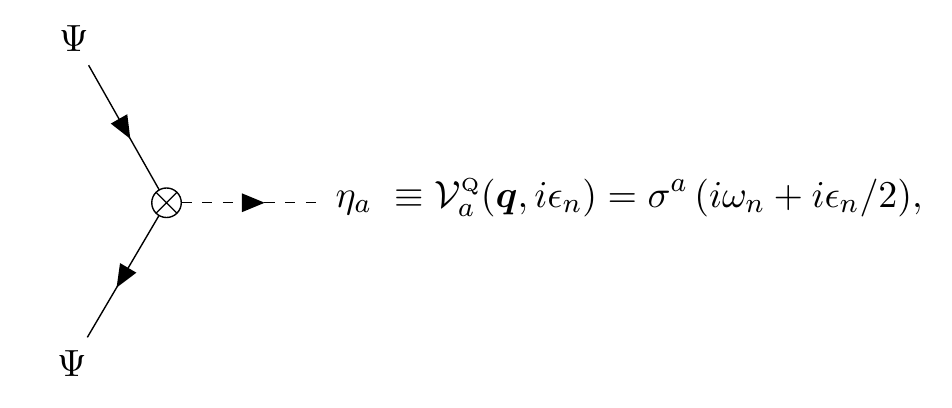}
\end{equation}
supplemented with a factor of $1/(\beta V)$ for each internal three-momentum. Likewise, recognizing that
\begin{equation}
\begin{split}
h (\vect{q}) &= \frac{1}{\sqrt{V}} e^{-i \vect{q}.\vect{r}_i} h^\pdagger_i \\&
= \frac{1}{2\sqrt{V}} \sum_{\vect{k}, \mu \nu} \left(h^{\mu \nu}_{\vect{k}+ \vect{q}} \psi^\dagger_{\vect{k}, \mu} \psi^\pdagger_{\vect{k}+\vect{q}, \nu} + h^{\nu \mu}_{\vect{k}} \psi^\dagger_{\vect{k}, \nu} \psi^\pdagger_{\vect{k}+\vect{q}, \mu} \right),
\end{split}
\end{equation}
we have, for $\hat{K}_{\vect{q}}$ as defined in Eq.~\eqref{eq:rel}, the second vertex:
\begin{equation}
\includegraphics[width=\linewidth]{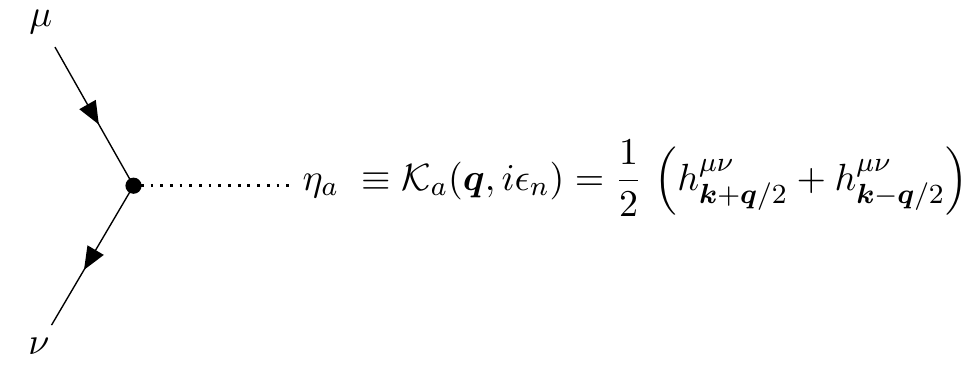}
\end{equation}
On top, for each independent momentum, a factor $1/(\beta \,V)$ remains from the corresponding Fourier transform.

Lastly, since there is no specific advantage in using the conventional relativistically invariant notation of Eq.~\eqref{eq:L_CS} at $T\ne0$, in some convenient situation, we use the following equivalent form of the fermion action
\begin{alignat}{1}
\label{eq:action}
\mathcal{L}&= \Psi_\alpha^{\dagger} \left(\partial^{\pdagger}_\tau - i A^{\pdagger}_\tau  -i \vect{\sigma}\cdot (\nabla - i \vect{A} )
      \right) \Psi^{\pdagger}_\alpha \nonumber\\
      &+  m\, \Psi_\alpha^{\dagger} \, \sigma^z\, \Psi_\alpha^{\pdagger}.
 \end{alignat}
Then, it is not difficult to see that propagator for the fermions is simply
\begin{equation}
\label{eq:psi_prop}
\feynmandiagram [horizontal=a to b] {a -- [fermion] b}; \equiv G^{\pdagger}_\Psi (\vect{k}, i\omega_n) =
 \frac{1}{-i\, \omega_n + \vect{\sigma} \cdot \vect{k} + m\, \sigma^z}.
\end{equation}
The action and propagator for the light and heavy fermions are exactly analogous, up to an appropriate substitution of $m_\ell$ or $M$ for $m$ in Eq.~\eqref{eq:psi_prop}.

\subsection{Noether procedure}

  Since, most of the time, we are dealing with a continuum field theory, it is also beneficial to directly write down the energy current or stress tensor of the field theory. In this section, we will use the Noether procedure to derive the stress tensor.

   First, we shall point out that our situation is different from the standard relativistic field theory because the spacetime is not Lorentzian. For example, the Dirac-Chern-Simons theory \eqref{eq:L_CS} comes from the underlying lattice model in Appendix.~\ref{LowEnergyFieldTheory}, and the gamma matrices in \eqref{eq:L_CS} acts on band index instead of physical spin index. Therefore, under spacetime rotation the fermions transform as spinless fields, and we have explicitly broken spin-statistics relation. The consequence of non-Lorentzian spacetime is that it is not always possible to covariantly couple the theory to conventional Riemannian metric and the stress tensor does not have to be symmetric. While it is possible to couple the theory to Newton-Cartan or Bargmann spacetime \cite{Geracie_2017}, we shall derive the stress-tensor in a simpler approach by using the Noether procedure.

  Next, we review the conventional Noether procedure for the stress tensor. We first apply a gauged spacetime translation $x^\mu\to x^\mu+\varepsilon^\mu(x)$ to the system. Because the system is translation invariant, the leading order response to $\varepsilon$ should be $\partial\varepsilon$, and the coefficient is defined to be the stress tensor:
  \begin{equation}\label{}
    \delta S=-\int \rd^3 x T^{\mu}_{~\nu}\partial_\mu\varepsilon^\nu.
  \end{equation}

  If we assume all fields transform as $\phi_a(x)\to\phi_a'(x)=\phi_a(x)-\varepsilon^\mu\partial_\mu \phi_a(x)$, we get
  \begin{equation}\label{eq:Tmunu_wrong}
    (T^\mu_{~\nu})_{incor.}=\frac{\partial\mathcal{L}}{\partial(\partial_\mu\phi_a)}\partial_\nu\phi_a-\delta^\mu_\nu\mathcal{L}.
  \end{equation}Here $\phi_a$ denotes all the field contents of the theory.

  The above formalism needs further improvement, because it does not respect gauge invariance, as can be seen by applying \eqref{eq:Tmunu_wrong} to a Maxwell theory. The conventional Belinfante improvement (see, for example, \cite{DiFrancesco:CFTbook}) is not applicable because it requires Lorentz symmetry. The spacetime-independent improvement is pointed out in \cite{Brauner}, by requiring the gauge field to transform correctly as a one-form:
  \begin{equation}\label{eq:deltaa}
       \delta a^a_\mu=-\varepsilon^\nu\partial_\nu a^a_\mu-\partial_\mu \varepsilon^\nu a^a_\nu.
   \end{equation} The additional term only depends on $\partial\varepsilon$, so it does not modify the global symmetry.

   The improved stress tensor is therefore
   \begin{equation}\label{}
     T^\mu_{~\nu}=(T^\mu_{~\nu})_{incor.}+\left(\frac{\partial\mathcal{L}}{\partial a^a_\mu}-\partial_\alpha\frac{\partial\mathcal{L}}{\partial(\partial_\alpha a^a_\mu)}\right)a^a_\nu.
   \end{equation}

  Applying the above formalism to the Dirac-Chern-Simons theoy \eqref{eq:L_CS}, we get
  \begin{equation}\label{eq:T_dirac}
    T^\mu_{~\nu}=\frac{1}{2}\bar{\Psi}_l i \gamma^\mu (\overrightarrow{\partial}_\nu-\overleftarrow{\partial}_\nu)\Psi_l+\bar{\Psi}_l\gamma^\mu a_\nu \bar{\Psi}_l-\delta^\mu_\nu\mathcal{L}_\Psi.
  \end{equation}

  For the Maxwell-Chern-Simons theory \eqref{MCS}, with $K_1$ and $K_2$ being constant, we get
  \begin{equation}\label{eq:T_MCS}
    T^{\mu\nu}=-\bar{f}^{\mu\rho}f_{~\rho}^{\nu}+\frac{1}{4}\eta^{\mu\nu}\bar{f}_{\alpha\beta}f^{\alpha\beta},
\end{equation}
  where $\bar{f}_{0i}=-\bar{f}_{i0}=K_2 f_{0i}$ and $\bar{f}_{ij}=-\bar{f}_{ji}=K_1 f_{ij}$. For the special case $K_1=K_2=1/g$, we get back to the standard result
  \begin{equation}\label{eq:T_K1=K2}
    T^{\mu\nu}=\frac{-1}{g}\left[f^{\mu\rho}f^{\nu}_{~\rho}-\frac{1}{4}\eta^{\mu\nu}f_{\alpha\beta}f^{\alpha\beta}\right].
  \end{equation}

\section{Gluon propagator}
\label{app:gluon}

The first component required to stitch together the diagram \ref{fig:gauge} is the gauge boson propagator. In this subsection we extend the calculations of Ref.~\onlinecite{kaul2008quantum} to the case of \textit{massive} fermions interacting with a SU(2) gauge field. The general structure of the gluon's effective action at large-$N$ follows from the Ward identity and is given by
\begin{eqnarray}
\label{eq:SeffTfin}
&\mathcal{S}^\pdagger_A = \frac{T}{2} \sum_{\epsilon_n} \int \frac{\mathrm{d}^2 \vect{q}}{4 \pi^2}
\Bigg[ \left( q^\pdagger_i A^\pdagger_\tau - \epsilon^\pdagger_n A^\pdagger_i \right)^2 \frac{D^\pdagger_{1} (\vect{q},
\epsilon_n)}{\vect{q}^2} \nonumber\\
&+ A^\pdagger_i A^\pdagger_j \left( \delta_{ij} - \frac{q^\pdagger_i q^\pdagger_j}{\vect{q}^2}
\right) D^\pdagger_{2} (\vect{q}, \epsilon_n) \Bigg],
\end{eqnarray}
where $D_1$ and $D_2$ are functions that can be evaluated at large-$N$ by perturbatively integrating out both the fermions starting from the action $\mathcal{S}$ in Eq.~\eqref{eq:action}. In Coulomb gauge $q_i A_i=0$, this yields the nonzero elements of the propagator to be
\begin{eqnarray}
 D^{ab}_{00}(\vect{q},\epsilon_n)&=&\frac{4\, \delta^{ab}}{D^\pdagger_1(\vect{q},\epsilon_n)},\\ \nonumber
D^{ab}_{ij}(\vect{q},\epsilon_n)&=&\left( \delta_{ij} - \frac{q_i q_j}{\vect{q}^2} \right) \frac{4\, \delta^{ab}}{D^\pdagger_2(\vect{q},\epsilon_n) +
(\epsilon_n^2/\vect{q}^2) D^\pdagger_1(\vect{q},\epsilon_n)},
\label{eq:propnzeroT1}
\end{eqnarray}
where $i,j$ run over the spatial indices only. The matrix structure in color space comes from inverting the product of Pauli matrices associated with the fermion loops in the functions $D_{1,2}$ as $\sum_{st} \tau^a_{st} \tau^b_{ts} = \mbox{Tr}\, (\tau^a \tau^b) = \delta^{ab}/4$. No fermions remain in the action $\mathcal{S}_A$ and all their effects are encapsulated in Eq.~\eqref{eq:SeffTfin} through these two functions alone.

Let us begin by calculating $D_1$, which is defined as:
\begin{alignat}{1}
\nonumber &D^\pdagger_1(\vect{q}, \epsilon_n) = - N  \int_{{\vect{k}},\omega_n}
\mbox{Tr} \left[
\mathcal{G}({\vect{k}},\omega_n)\,\mathcal{G}({\vect{q}+\vect{k}},\omega_n + \epsilon_n)
\right]\\
&  = 2 N \int_{{\vect{k}},\omega_n}
\frac{\omega_n
(\omega_n+\epsilon_n) - m^2- {\vect{k}} \cdot ({\vect{k}} + \vect{q})}{(\omega_n^2 + \vect{k}^2 + m^2)((\omega_n + \epsilon_n)^2 + ({\vect{k}} +
{\vect{q}})^2 + m^2) },
\end{alignat}
where we use the shorthand $\int_{{\vect{k}},\omega_n} = T \sum_{\omega_n}
\int \mathrm{d}^2 \vect{k}/(4 \pi^2)$ to signify a summation on the internal
frequencies and momenta. Using the Passarino-Veltman reduction formula \cite{passarino1979one}, this can be manipulated into
\begin{alignat}{1}
\label{eq:twoeq}
&D^\pdagger_1(\vect{q}, \epsilon_n) = N \int_{{\vect{k}},\omega_n}
\Bigg[\frac{-2}{\omega_n^2 + \vect{k}^2 + m^2} \nonumber\\&+  \frac{(2 \omega_n +\epsilon_n)^2
+\vect{q}^2}{(\omega_n^2 + \vect{k}^2+m^2)((\omega_n + \epsilon_n)^2 + ({\vect{k}} + {\vect{q}})^2 + m^2)} \Bigg].
\end{alignat}
For the first of the two integrals here, the UV divergence is linear, so it is most convenient to use $\zeta$-function regularization \cite{shiekh1990zeta} in which
\begin{equation}
\int_0^\infty \mathrm{d} x = 0, \quad \int_1^\infty \frac{1}{x}\, \mathrm{d} x  = \mbox{ arbitrary}.
\end{equation}
Then, within this scheme,
\begin{alignat}{1}
\nonumber &~N \int_{{\vect{k}},\omega_n} \frac{-2}{\omega_n^2 + \vect{k}^2 + m^2} \\
\nonumber&= -N \int \frac{\mathrm{d}^2 \vect{k}}{4 \pi^2} \frac{\tanh \left(\frac{1}{2} \beta  \sqrt{\vect{k}^2+m^2}\right)}{\sqrt{\vect{k}^2+m^2}} \\&
\nonumber= \frac{N}{2 \pi} \int \mathrm{d} k \left(1 -   \frac{k \tanh \left(\frac{1}{2} \beta  \sqrt{k^2+m^2}\right)}{\sqrt{k^2+m^2}} \right)\\
&= \frac{N \,T}{\pi}\left[ \ln 2 + \ln \left(\cosh \left(\frac{\beta\, \lvert m \rvert}{2}\right)\right)\right].
\end{alignat}
The second integral in Eq.~\eqref{eq:twoeq} can be evaluated by introducing Feynman parameters and shifting the loop momentum $\vect {k} \rightarrow \vect{k} - u \vect{q}$:
\begin{widetext}
\begin{alignat}{1}
&N \int_0^1 \mathrm{d} u \int_{{\vect{k}},\omega_n} \frac{(2 \omega_n +\epsilon_n)^2
+\vect{q}^2}{\left[u \,(\omega_n + \epsilon_n)^2 + (1-u)\,\omega_n^2   +  ({\vect{k}} + u {\vect{q}})^2 + u (1-u) \,\vect{q}^2 + m^2\right]^2}\\
\nonumber= &\frac{N \,T}{4\pi} \sum_{\omega_n} \left[(2 \omega_n +\epsilon_n)^2
+\vect{q}^2\right] I^{(0)}_{n}; \mbox{ with } I^{(0)}_{n} = \int_0^1 \mathrm{d} u  \frac{1}{u \,(\omega_n + \epsilon_n)^2 + (1-u)\,\omega_n^2   +  u (1-u) \,\vect{q}^2 + m^2}\\
= &\frac{N \,T}{4\pi} \sum_{\omega_n} \bigg\{ \left((2 \omega_n +\epsilon_n)^2
+\vect{q}^2\right)\,\frac{1}{\mathcal{A}_n}\ln \left(\displaystyle \frac{2 m^2+\vect{q}^2+2 \omega_n ^2+\epsilon_n ^2+2 \omega_n  \epsilon_n +\mathcal{A}_n}{ 2 m^2+\vect{q}^2+2 \omega_n ^2+\epsilon_n ^2+2 \omega_n  \epsilon_n -\mathcal{A}_n}\right) \bigg\},
\label{eq:curly1}
\end{alignat}
\end{widetext}
where $\mathcal{A}_n \equiv \sqrt{4 m^2 \vect{q}^2+\left( \vect{q}^2+\epsilon_n^2\right) \left( \vect{q}^2+(2 \omega_n +\epsilon_n )^2\right)} $. We shall encounter the integral $I^{(0)}_{n}$ in multiple contexts later so it is handy to define it separately. The summation over the fermionic Matsubara frequencies can only be performed numerically, and in this regard, it is useful to establish the large-$\omega_n$ behavior of the terms since $\omega_n$ is not bounded above. After symmetrizing over positive and negative frequencies and subtracting the divergent piece using a $\zeta$ regulator, the $1/\omega_n$ expansion for the terms within the curly braces in Eq.~\eqref{eq:curly1} stands as
\begin{widetext}
\begin{alignat}{1}
\label{eq:D1int}
&\frac{-12 m^2+\vect{q}^2+\epsilon_n ^2}{3\, \omega_n ^2} + \frac{120\, m^4+10 m^2 \left(\vect{q}^2-15 \epsilon_n ^2\right)-\vect{q}^4+8 \vect{q}^2 \epsilon_n ^2+9 \epsilon_n ^4}{30\, \omega_n ^4}\\
\nonumber &+ \frac{1}{\omega_n ^6} \left [\frac{1}{210} \left(\vect{q}^2+\epsilon_n ^2\right) \left(\vect{q}^4-19 \vect{q}^2 \epsilon_n ^2+50 \epsilon_n ^4\right)-4 m^6- m^4 \left(\vect{q}^2-15 \epsilon_n ^2\right)-\frac{1}{15} m^2 \left(\vect{q}^4-14 \vect{q}^2 \epsilon_n ^2+105 \epsilon_n ^4\right) \right]
\end{alignat}
\end{widetext}
followed by terms of $\mathcal{O}(1/\omega_n^8)$. Note that Eq.~\eqref{eq:D1int} reduces correctly to the expressions documented in Ref.~\cite{kaul2008quantum} in the limit $m \rightarrow 0$. This asymptotic behavior can then be summed by using the identities
\begin{eqnarray}
\sum_{n=M+1}^{\infty} \frac{1}{(2n-1)^2} &=& \frac{1}{4M} -
\frac{1}{48M^3} + \frac{7}{960 M^5} + \ldots, \nonumber \\
\sum_{n=M+1}^{\infty} \frac{1}{(2n-1)^4} &=&
\frac{1}{48M^3} - \frac{1}{96 M^5} + \ldots, \nonumber \\
\sum_{n=M+1}^{\infty} \frac{1}{(2n-1)^6} &=& \frac{1}{320 M^5} +
\ldots,
\end{eqnarray}
while retaining the exact functional dependence of Eq.~\eqref{eq:curly1} for small $\omega_n$ up to $2 \pi\, (\pm M +1/2) T$.

We can employ a similar procedure for the second function in Eq.~\eqref{eq:propnzeroT1}:
\begin{widetext}

\begin{alignat}{1}
D^\pdagger_2(\vect{q}, \epsilon_n) &= - \frac{\vect{q}^2}{q_x \,q_y} N  \int_{{\vect{k}},\omega_n}
\mbox{Tr} \left[ \sigma^x\,
\mathcal{G}({\vect{k}},\omega_n)\,\sigma^y\,\mathcal{G}({\vect{q}+\vect{k}},\omega_n + \epsilon_n)
\right]\\
\nonumber &= - \frac{2\,  \vect{q}^2}{q_x \,q_y} N  \int_{{\vect{k}},\omega_n}
\frac{2 k_x k_y + k_x q_y + k_y q_x + m\, \epsilon_n}{(\omega_n^2 + \vect{k}^2 + m^2)((\omega_n + \epsilon_n)^2 + ({\vect{k}} +
{\vect{q}})^2 + m^2) }\\
\nonumber &=  \frac{N \,T}{2 \pi } \frac{\vect{q}^2}{q_x \,q_y}\sum_{\omega_n}  \int_0^1 \mathrm{d} u \frac{2\, q_x\, q_y\, u\, (1-u) - m\,\epsilon_n}{u \,(\omega_n + \epsilon_n)^2 + (1-u)\,\omega_n^2  + u (1-u) \,\vect{q}^2 + m^2}\\
&=  \frac{N \,T}{2 \pi } \frac{\vect{q}^2}{q^\pdagger_x \,q^\pdagger_y}\sum_{\omega_n}
\left[2\, q^\pdagger_x\, q^\pdagger_y\, I^{(2)}_n - m\,\epsilon^\pdagger_n I^{(0)}_n \right],
\label{eq:D2_part1}
\end{alignat}
where $I^{(0)}_{n}$ has already been calculated earlier and $I^{(2)}_{n}$ is defined as the integral
\begin{alignat}{1}
I^{(2)}_{n} &= \int_0^1 \mathrm{d} u  \frac{u\, (1-u)}{u \,(\omega_n + \epsilon_n)^2 + (1-u)\,\omega_n^2   +  u (1-u) \,\vect{q}^2 + m^2},\\
\nonumber&=\bigg[\left(2 \vect{q}^2 \left(m^2+\omega_n ^2\right)+\epsilon_n  (2 \omega_n +\epsilon_n ) \left(\vect{q}^2+ 2\epsilon_n \omega_n +\epsilon_n^2 \right)\right) \ln \left(\frac{(\mathcal{C}_n  -\vect{q}^2)^2-(\epsilon_n ^2+2 \omega_n  \epsilon_n)^2}{(\mathcal{C}_n  +\vect{q}^2)^2-(\epsilon_n ^2+2 \omega_n  \epsilon_n)^2}  \right)\\
&\qquad+\mathcal{C}_n   \left(\epsilon_n  (2 \omega_n +\epsilon_n ) \ln \left(\frac{m^2+(\omega_n +\epsilon_n )^2}{m^2+\omega_n ^2}\right)+2 \vect{q}^2\right) \bigg]  \frac{1}{2\, \mathcal{C}_n  \vect{q}^4}
\end{alignat}
with $\mathcal{C}_n  \equiv \sqrt{4 \vect{q}^2 \left(m^2+\omega_n^2\right)+\left(\vect{q}^2+\epsilon_n (2 \omega_n +\epsilon_n )\right)^2}$. The full expression for  Eq.~\eqref{eq:D2_part1} is forbiddingly complex and is also not particularly insightful. Instead, akin to  Eq.~\eqref{eq:D1int}, we can symmetrize over the frequencies and write out $D_2$ for large $\omega_n$ in a series expansion as
\begin{alignat}{2}
&\nonumber D^\pdagger_2(\vect{q}, \epsilon_n) &&= \frac{N \,T\,\vect{q}^2 }{2 \pi }\sum_{\omega_n} \bigg[  \frac{q_x q_y-3 m \epsilon_n }{3\, q_x\, q_y\,\omega^2_n} +\frac{5 m \epsilon_n  \left(6 m^2+\vect{q}^2\right)-2 q_x q_y \left(5 m^2+\vect{q}^2\right)-25 m \epsilon_n ^3+7 q_x q_y \epsilon_n ^2}{30\, q_x\, q_y\,\omega_n^4}\\
\nonumber& &&+\frac{1}{210\, q_x\, q_y \,\omega_n^6} \bigg\{\big(210 m^5 \epsilon_n -70 m^4 q_x q_y+70 m^3 \epsilon_n  \left(\vect{q}^2-9 \epsilon_n ^2\right)-14 m^2 q_x q_y \left(2 \vect{q}^2-13 \epsilon_n ^2\right)\\
& &&+7 m \epsilon_n  \left(-13 \epsilon_n ^2 \vect{q}^2+\vect{q}^4+16 \epsilon_n ^4\right)+q_x q_y \left(34 \epsilon_n ^2 \vect{q}^2 -3 \vect{q}^4-13 \epsilon_n ^4\right)\big) \bigg\} + \mathcal{O}\left(\frac{1}{\omega_n^8} \right)\bigg],
\end{alignat}
which is convergent at large $\omega_n$. Once again, when $m = 0$, this correctly reproduces the results of \citet{kaul2008quantum}.

\subsection{Limit of zero external momentum}

While the propagator derived above holds for all momenta, the $\vect{q}=0$ limit, in particular, involves some subtleties and must be dealt with care. In the limit where the external momentum is zero, $D_1$ is finite, and according to Eq.~\eqref{eq:curly1}, goes to
\begin{equation}
\frac{N \,T}{4\pi} \sum_{\omega_n}\frac{\lvert2 \omega_n +\epsilon_n \rvert}{\lvert \epsilon_n \rvert}  \ln \left(\displaystyle \frac{2 m^2+\omega_n ^2+ (\epsilon_n + \omega_n )^2 +\lvert \epsilon_n\rvert\lvert2 \omega_n +\epsilon_n \rvert}{2 m^2+\omega_n ^2+ (\epsilon_n + \omega_n )^2 -\lvert \epsilon_n\rvert\lvert2 \omega_n +\epsilon_n \rvert}\right),
\label{eq:D10q}
\end{equation}
so the temporal component of the gluon propagator, $D_{00}^{ab}$, is nonzero. The more nontrivial part is the spatial component
\begin{alignat}{1}
D_{ij}(\vect{q},\epsilon_n)&=\frac{\vect{q}^2 \delta_{ij} -q_i\, q_j  }{\vect{q}^2\,D^\pdagger_2(\vect{q},\epsilon_n) +
\epsilon_n^2 D^\pdagger_1(\vect{q},\epsilon_n)} ,
\end{alignat}
and specifically, the behavior of $D_2 (\vect{q},\epsilon_n)$ as $\vect{q} \rightarrow 0$:
\begin{alignat}{1}
\lim_{\vect{q} \rightarrow 0} \vect{q}^2\,D^\pdagger_2(\vect{q},\epsilon_n) &= \frac{N \,T}{2 \pi } \sum_{\omega_n} \bigg[\epsilon_n  (2 \omega_n +\epsilon_n ) \ln \left(\frac{m^2+(\omega_n +\epsilon_n )^2}{m^2+\omega_n ^2}\right) \\
\nonumber&- \frac{2 m\, \vect{q}^4 \epsilon_n}{q_x q_y\lvert \epsilon_n \rvert \lvert 2 \omega_n + \epsilon_n \rvert} \left \{ \tanh ^{-1}\left(\frac{\epsilon_n ^2+2 \omega_n  \epsilon_n }{\lvert \epsilon_n \rvert \lvert 2 \omega_n + \epsilon_n \rvert}\right)+ \tanh ^{-1}\left(\frac{\epsilon_n ^2-2 \omega_n  \epsilon_n }{\lvert \epsilon_n \rvert \lvert 2 \omega_n + \epsilon_n \rvert}\right)\right\} \bigg].
\end{alignat}
Rewriting this in polar coordinates, and assuming $\cos, \sin \theta \ne 0$, we find
\begin{alignat}{1}
\nonumber D_{ij}(q, \theta,\epsilon_n)&\sim\frac{q^2 \delta_{ij} -q^2 (\cos^{4-i-j} \theta \,\sin^{i+j-2} \theta) }{\displaystyle \frac{q^4}{q^2 \cos \theta \sin \theta} \,\chi^\pdagger_0 (\epsilon^\pdagger_n)+ \chi^\pdagger_1 (\epsilon^\pdagger_n)} = \frac{q^4 \left(\delta_{ij} - \cos^{4-i-j} \theta \,\sin^{i+j-2} \theta \right) \cos \theta \sin \theta}{\displaystyle q^4 \,\chi^\pdagger_0 (\epsilon^\pdagger_n)+ q^2 \cos \theta \sin \theta\, \chi^\pdagger_1 (\epsilon^\pdagger_n)} = 0,
\end{alignat}
\end{widetext}
where the $ \chi$ are functions of $\epsilon_n$ alone, independent of $\vect{q}$. Thus, the spatial components of the gluon propagator are zero when the external momentum is zero. The same result can be proved even when the assumption above is relaxed by successively taking the limits $ q_x \rightarrow 0,\,  q_y \rightarrow 0$.

\section{Framing anomaly in the `wrong' metric}
\label{app:wrong}

In the main text, we have calculated the framing anomaly using a metric compatible with the speed of `light'  $c_0=\sqrt{K_1/K_2}$. It would also be interesting to put the theory in an incompatible metric whose speed of `light' is different from $c_0$ and redo the computation. We expect the result to be essentially the same as the one obtained from a compatible metric.

Let us assume we are in a spacetime with metric $\eta_{\mu\nu}=(1,-1,-1)$, and the MCS theory \eqref{MCS} has a speed of `light' $c_0=\sqrt{K_1/K_2}\neq 1$.

In momentum space, the MCS theory has the following form (we have included a gauge-fixing term)
  \begin{equation}\label{eq:S=aKa}
  \begin{split}
    &S=\int\frac{\rd^3 p}{(2\pi)^3}\frac{K_2}{2}a_\mu(-p)a_\nu(p)\\&\times
    \left(c_0^2 P_1^{\mu\nu}(p)+P_2^{\mu\nu}(p)+\frac{p^\mu p^\nu}{\xi}+m_t\varepsilon^{\mu\nu\rho}ip_\rho\right).
  \end{split}
  \end{equation} Here $P_1$, $P_2$ are the transverse projectors corresponding to $\vect{B}^2,~\vect{E}^2$ respectively:
  \begin{eqnarray}
    P_1^{\mu\nu} &=& \begin{pmatrix}
                       0 & 0 \\
                       0 & p^i p^j-\delta^{ij}\vect{p}^2
                     \end{pmatrix}, \\
    P_2^{\mu\nu} &=& \begin{pmatrix}
                       \vect{p}^2 & p^0p^j \\
                       p^0p^i & (p^0)^2\delta^{ij}
                     \end{pmatrix}.
  \end{eqnarray}

Inverting the matrix in the parenthesis of Eq.~\eqref{eq:S=aKa}, we get the gauge field propagator (in $\xi=0$ gauge)
   \begin{equation}\label{}
     D^{\mu\nu}(p)=A_1 P_1^{\mu\nu}+A_2 P_2^{\mu\nu}+A_3 \varepsilon^{\mu\nu\rho}ip_\rho,
   \end{equation}where
   \begin{eqnarray}
     A_1  &=& \frac{1}{K_2p^2(\tilde{p}^2-m_t^2)}\frac{(2-c_0^2)(p^0)^2-\vect{p}^2}{p^2}, \\
     A_2 &=& \frac{1}{K_2p^2(\tilde{p}^2-m_t^2)}\frac{\tilde{p}^2}{p^2}, \\
     A_3 &=& \frac{-m_t}{K_2p^2(\tilde{p}^2-m_t^2)},
   \end{eqnarray} and $\tilde{p}^2=(p^0)^2-c_0^2\, \vect{p}^2$.

   Next, we discuss the stress tensor $T^{\mu\nu}$. Since $K_1\neq K_2$, there is no natural way to couple the system to a background metric, so we have to use Noether's theorem to derive $T^{\mu\nu}$. To ensure gauge invariance, we use a modified Noether procedure which is described in Appendix.~\ref{app:feynmanrules}. Using the transformation law \eqref{eq:deltaa}, we can write down the stress tensor
\begin{equation}
    T^{\mu\nu}=-\bar{f}^{\mu\rho}f_{~\rho}^{\nu}+\frac{1}{4}\eta^{\mu\nu}\bar{f}_{\alpha\beta}f^{\alpha\beta},
\end{equation}
  where $\bar{f}_{0i}=-\bar{f}_{i0}=K_2 f_{0i}$ and $\bar{f}_{ij}=-\bar{f}_{ji}=K_1 f_{ij}$. This result agrees with the energy-momentum tensor of classical electrodynamics in a medium.

   The computation of the gravitational Chern-Simons term and the thermal Hall effect can now be carried out in the same way as in the main text. In this calculation, the cancellation of the $p^2$ factors seen in \eqref{eq:Pi_AS=u} also happens.
   Therefore, the denominator of the integrand is now $(\tilde{p}^2-m_t^2)((\tilde{p}+\tilde{q})^2-m_t^2)$.
   At zero temperature, the momentum integral can be performed in standard ways after rescaling the zeroth component, yielding the following gravitational Chern Simons term:
    \begin{equation}\label{eq:CSgh2}
    \begin{split}
        CS_g[h]=\frac{-c}{192\pi}\int\frac{\rd^3p}{(2\pi)^3}h_{\mu\nu}(-p)\varepsilon^{\mu\rho\sigma}(ip_\sigma)\tilde{P}_T^{\nu\lambda}h_{\rho\lambda}(p),
    \end{split}
   \end{equation} where $\tilde{P}_T$ is a transverse projector in the compatible metric:
   \begin{equation}
    \begin{split}
       \tilde{\eta}_{\mu\nu}&=(c_0^2,-1,-1),\\
       \tilde{P}_{T{\mu\nu}}&=\tilde{\eta}_{\mu\nu}\frac{\tilde{p}^2}{c_0^2}-p_\mu p_\nu.
    \end{split}
   \end{equation} A subtlety here is that all indices are raised and lowered with the incompatible metric $\eta_{\mu\nu}=(1,-1,-1)$.

   As for the thermal hall effect, we compute the antisymmetrized polarization analogous to \eqref{eq:Pi_AS=u}, which now becomes
   \begin{equation}\label{}
     \Pi_{\text{AS}}^{\mu0;\rho0}=\int\frac{\rd^3 q}{(2\pi)^3}\frac{-m_t\varepsilon^{\mu\rho\sigma}\tilde{u}_\sigma}{(\tilde{q}^2-m_t^2)((\tilde{p}+\tilde{q})^2-m_t^2)},
   \end{equation} and the $\tilde{u}_\sigma$'s are related to the $u_\sigma$'s in the main text by simple scaling:
   \begin{eqnarray}
     \tilde{u}^0(p^0,q^0,\vect{p},\vect{q}) &=& c_0^2 u^0(p^0,q^0,c_0\vect{p},c_0\vect{q}) \\
     \tilde{u}^i(p^0,q^0,\vect{p},\vect{q}) &=& c_0 u^i(p^0,q^0,c_0\vect{p},c_0\vect{q}).
   \end{eqnarray}

    Carrying out the integration, we found that \eqref{eq:Ap} is not altered, and therefore the thermal Hall coefficient remains to be
   \begin{equation}\label{eq:kappaxy2}
       \kappa^{}_{xy}=\frac{\pi}{6}c\,T.
   \end{equation}

\bibliography{TH}

\end{document}